\documentclass[lettersize,journal]{IEEEtran}
\usepackage{amsmath,amsfonts}
\usepackage{algorithm}
\usepackage{array}
\usepackage[caption=false,font=normalsize,labelfont=sf,textfont=sf]{subfig}
\usepackage{textcomp}
\usepackage{stfloats}
\usepackage{url}
\usepackage{verbatim}
\usepackage{graphicx}
\usepackage{cite}
\DeclareMathAlphabet\mathzapf       {T1}{pzc} {mb} {it}
\newcommand{\funC}{\mathzapf{C}}
\newcommand{\funF}{\mathzapf{F}}

\usepackage{algorithm}% http://ctan.org/pkg/algorithm
\usepackage{algpseudocode}% http://ctan.org/pkg/algorithmicx

\newcommand{\citep}[1]{\cite{#1}}
\begin{document}

\title{Deep Dynamic Cloud Lighting}

\author{
\IEEEauthorblockN{Pinar Satilmis\IEEEauthorrefmark{1}, Thomas Bashford-Rogers\IEEEauthorrefmark{2}} \\
\IEEEauthorblockA{\IEEEauthorrefmark{1}Birmingham City University}
\IEEEauthorblockA{\IEEEauthorrefmark{2}The University of Warwick}
}

\markboth{Journal of \LaTeX\ Class Files,~Vol.~14, No.~8, August~2021}%
{Shell \MakeLowercase{\textit{et al.}}: A Sample Article Using IEEEtran.cls for IEEE Journals}

\IEEEpubid{0000--0000/00\$00.00~\copyright~2021 IEEE}

\maketitle

\begin{abstract}

Sky illumination is a core source of lighting in rendering, and a substantial amount of work has been developed to simulate lighting from clear skies. However, in reality, clouds substantially alter the appearance of the sky and subsequently change the scene illumination. While there have been recent advances in developing sky models which include clouds, these all neglect cloud movement which is a crucial component of cloudy sky appearance. In any sort of video or interactive environment, it can be expected that clouds will move, sometimes quite substantially in a short period of time. Our work proposes a solution to this which enables whole-sky dynamic cloud synthesis for the first time. We achieve this by proposing a multi-timescale sky appearance model which learns to predict the sky illumination over various timescales, and can be used to add dynamism to previous static, cloudy sky lighting approaches.
\end{abstract}


\section{Introduction} \label{Sec:Introduction}

The sky is a vital illumination source when rendering 3D scenes. A core component of real skies is clouds which can substantially alter the scene illumination compared to clear sky lighting, and as such, need to be represented when generating imagery. This typically takes the form of whole-sky illumination, which represents lighting information for the whole hemisphere of the sky above the scene.

Sky illumination can be computed by using environment maps, analytical models or by simulating the radiative transport equation. Simulating the radiative transport equation can generate highly realistic clouds. However, this is computationally demanding and requires 3D volumetric representations of complicated cloud structures which are challenging to model or generate. Environment maps are High Dynamic Range (HDR) images which store far-field directional illumination and are used for image-based lighting; see Debevec \cite{debevec2008rendering}. These typically are captured using specialized devices which capture 360$^{o}$ images or using fish-eye images to capture the upper hemisphere containing sky illumination. Though this method can represent lighting from cloudy skies, these methods are limited by the capturing process to a fixed number of locations, times, and cloud types present during capture.  Analytical models can reproduce a wide range of sky scenarios and create dynamic lighting for different solar positions, \cite{hosek2012analytic, satilmis2016machine, wilkie2021fitted}. However, these methods represent clear sky lighting and do not consider clouds. 

To overcome these limitations, recently generative machine learning methods have been developed \cite{satilmis2022deep, mirbauer2022skygan} to generate cloudy sky environment maps. These typically are trained on images of real cloudy skies and are combined with an analytical sky model input to generate clouds either via a Generative Adversarial Network (GAN) as in the work by Mirbauer et al. \cite{mirbauer2022skygan}, or via user or artist specified masks and a U-net as  used in Satilmis et al. \cite{satilmis2022deep}. These methods can generate realistic images of cloudy skies suitable for use as environment maps without the complexities of representing and rendering cloud volumes.

\begin{figure*}
    \centering
    \includegraphics[scale=0.35]{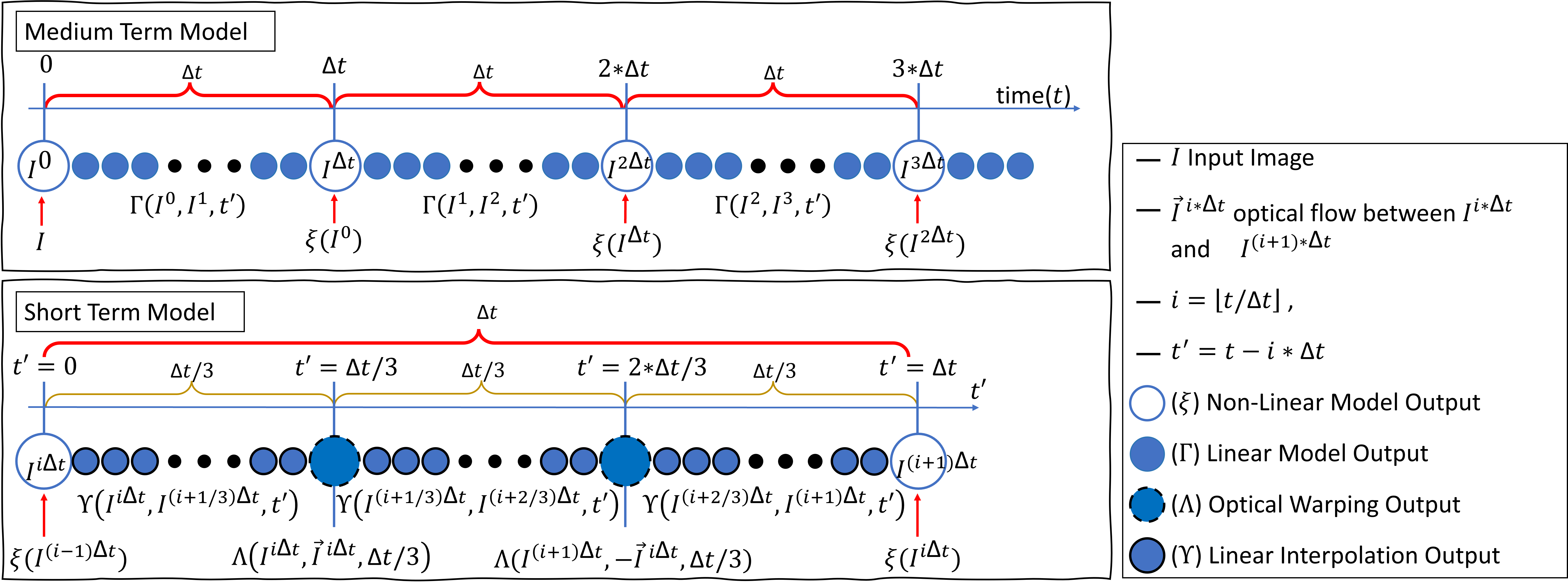}
    \caption{An overview of deep dynamic cloud generation method. The method uses a non-linear model ($\xi$) and a linear model ($\Gamma$) to predict the sky lighting from a  given input cloudy sky-image ($I$). $\xi$ predicts the sky-image from its previous prediction for $\Delta t$ time interval. $\Gamma$ interpolates the sky lighting in between two sequential predictions of $\xi$. $\Gamma$ is optical warping for $t' \in \{\Delta t/3,2\Delta t/3\}$, and is linear interpolation for $t' \in (0, \Delta t) \setminus \{\Delta t/3,2\Delta t/3\}$.}
    \label{fig:overview}
\end{figure*}

Despite the ability of these generative methods to synthesize cloudy sky imagery, these aforementioned methods still lack one important feature of clouds: dynamism. Clouds are not static and can significantly change their position in a relatively short period of time. Existing methods focus on the generation of a single frame of cloudy sky illumination, but when used in practice, the dynamism of cloud movement needs to be included in a generative model. Examples of this use case are clouds moving in a rendered environment for an animated sequence for film, or used to provide dynamic skies in interactive entertainment applications. This work proposes an approach which can synthesize plausible cloud movement given a single input image, either captured or generated by a generative machine learning model.

We achieve this by proposing a multi-timescale approach to cloud synthesis. At a longer timescale, a deep learning model predicts the larger non-linear changes in cloud positions and associated appearance. It achieves this via neural networks which predict a flow field of how clouds move across the sky, and based on this an illumination at the previous time step, predicts cloudy sky appearance at the next time step. Cloud movement at shorter timescales is predicted via a linear model of cloud movement conditioned on the nonlinear movement from the longer timescale and leads to the smooth movement of cloud position, shape, and illumination between the longer timesteps.

Our approach is trained and validated using a database of captured sequences of cloud movement, and we show how this can be applied to either a single captured cloudy environment image, or an image generated by a generative cloud model \cite{satilmis2022deep, mirbauer2022skygan}. Our technique allows animation of artistically generated clouds and produces results which can be directly used in a rendering system for environment illumination.

To summarize, the main contributions of the paper are:

\begin{itemize}
    \item A novel framework for synthesizing dynamic cloud lighting from a single input image, either from a hemispherical sky capture or from recent static generative methods.
    \item A multi-timescale approach which can generate longer timescale cloud movement while ensuring smooth, coherent movement at short timescales.
    \item Results demonstrating our approach can synthesize smooth cloud movement and show this applied in different rendering scenarios.
\end{itemize}

\newpage

\section{Related Work} \label{Sec:RelatedWork}

In this section, we cover the main approaches to generating sky illumination. First, we discuss clear sky models which provide cloud-free sky illumination; then we cover the two main approaches to synthesize sky illumination with clouds: simulating cloud structure, dynamics, and lighting via numerical simulations and using deep image-generation techniques to generate cloudy sky illumination.

\subsection{Clear Sky} \label{Subsec:ClearSky}

Clear sky models are typically based on analytical or tabulated approximations of atmospheric light transport. These models combine sky specifications, for example, turbidity, solar position, and ground albedo, and predict incoming light from a given direction. An early model used in graphics was proposed by Perez \cite{perez1993all}. When the parameters are carefully chosen, this five-parameter model can accurately describe low-turbidity skies. Later, models that were proposed based on fitting results to brute-force atmospheric light transport simulations, such as Nishita et al. \cite{nishita1993display, nishita1996display} Haber et al. \cite{haber2005physically}, Preetham \cite{preetham1999practical},  Hosek and Wilkie \cite{hosek2012analytic} and Wilkie et al. \cite{wilkie2021fitted}. These all increase the quality of the approximations of the clear sky illumination by including aspects such as multiple scattering in the atmosphere such as Hosek and Wilkie \cite{hosek2012analytic} and realistic profiles of scattering particles such as Wilkie et al. \cite{wilkie2021fitted}. These methods also typically trade quality for memory usage, where previous models used fewer parameters which led to a limited range of representable phenomena, more recent models have generated more accurate sky illumination at a large memory cost.

\subsection{Cloud Modeling} \label{Subsec:CloudModelling}

Modeling cloud structure is typically achieved via either procedural methods or physical simulation. Procedural methods include using implicit functions, Ebert \cite{ebert1997volumetric}; fractals, Voss \cite{voss1983fourier}; textured ellipsoids, Gardner \cite{gardner1985visual}; spectral models, Sakas \cite{sakas1993modeling}; and implicit ellipsoids, Schpok et al. \cite{schpok2003real}. Physically based models are derived from images of clouds and different approaches have been investigated. Dobashi et al. \cite{dobashi1999using} modeled clouds as metaballs from satellite images. The method by Wither et al. \cite{wither2008rapid} generated a cloud mesh from a user-drawn sketch. Dobashi et al. \cite{dobashi2010modeling} and Yuan et al. \cite{yuan2014modelling} focused on modeling clouds from a single input image.

Simulating clouds is also widely investigated in the literature, for example, Dobashi et al. \cite{dobashi2000simple}, Harris
and Lastra \cite{harris2001real}, and Dobashi et al. \cite{dobashi2008feedback}. Recently, the complex nature of cloud formation was investigated by H{\"a}drich et al. \citep{hadrich2020stormscapes} who developed a novel framework to simulate physically accurate clouds. This led to plausible cloud dynamics in 3D, but it is computationally expensive and furthermore needs to be coupled with expensive light transport simulation to produce realistic imagery. For more details about approaches to representing clouds in computer graphics, please see the survey by Goswami \cite{goswami2021survey}.

\subsection{Volumetric Cloud Rendering} \label{Subsec:CloudySkyLight}

Generating physically accurate cloud renderings is a computationally heavy task, especially if animations are required, or the clouds need to be synthesized into an environment map for use in conventional rendering software.

Motivated by this, several fast specialized, methods for cloud rendering have been proposed to generate plausible cloud appearance Harris and Lastra \cite{harris2001real}, Dobashi et al. \cite{dobashi2000simple}, Riley et al. \cite{riley2004efficient}, Nishita et al. \cite{nishita1996display}, Elek and Kmoch \cite{elek2010real}. When higher levels of realism are required, physically based light transport methods can be used. These simulate light transport through the clouds and atmosphere. This is commonly achieved by Monte-Carlo volumetric path tracing; see Nov{\'a}k et al. \citep{novak2018monte} for a detailed explanation. Despite being physically accurate, this rendering technique is very computationally intensive, especially for sky and cloud rendering which has a very large spatial volume combined with cloud scattering parameters which lead to hundreds of scattering events for each path. To accelerate this process, Kallweit et al. \cite{kallweit2017deep} applied neural networks to improve the efficiency of Monte-Carlo rendering. Though these methods can achieve highly realistic cloud imagery, this does not take into account the subtleties of light transport in real clouds, for example, non-exponential free-flight distributions. This was addressed by Bitterli et al. \citep{bitterli2018radiative} and Jarabo et al. \citep{Jarabo2018radiative} concurrently.

These methods can approximate cloudy sky illumination, but are very computationally expensive, and still can lead to non-photorealistic results. 

\subsection{Deep learning based methods} \label{Subsec:IBL}

To avoid the complexities of simulating light transport when generating environment maps for later use in rendering, several deep learning based approaches for generating cloudy sky illumination have been proposed. These directly generate pixels in an environment map given a specification of the sky and cloud conditions.

Satilmis et al. \citep{satilmis2022deep} and Mirbauer et al. \citep{mirbauer2022skygan} proposed methods to generate whole-sky cloudy lighting which is suitable for use as an environment map and therefore can be directly integrated into rendering systems. Satilmis et al. \citep{satilmis2022deep} used U-net structured autoencoders to generate cloudy sky lighting from encoded clear sky lighting and a cloud mask. Mirbauer et al. \citep{mirbauer2022skygan} used generative adversarial networks to synthesize cloudy skies conditioned on sun position and cloud coverage.

A different approach using transformers is developed by Chen et al. \cite{chen2022text2light} to generate HDR panoramas from a given text description. Goswami et al. \cite{goswami2023interactive} produced a method that synthesized cloud animations in world space but relied on simplified volumetric cloud and lighting models.

These methods propose photorealistic sky lighting which includes clouds, however are limited to generating one frame of illumination. As none of these approaches consider the temporal aspect of sky illumination, they cannot produce a coherent series of frames for cloud animation. Our work lifts this limitation.

\subsection{Deep image synthesis and editing} \label{Subsec:DeepISE}

There has also been work on generating or editing images with clouds in image space. This is different from our application as these approaches manipulate clouds in a perspective view, whereas we are focused on the full hemisphere of lighting. These approaches often utilize a semantic layout, for example, Chen and Koltun \cite{chen2017photographic} used a single feedforward network to synthesize photographic images. Although it can generate high-resolution images, it lacks the high-frequency details needed to represent realistic clouds. Park et al. \cite{park2019semantic} used GANs to synthesize images with skies and clouds, enabling both semantic and style user control; see Tewari et al. \cite{tewari2020state} and Tewari et al. \cite{tewari2022advances} for a wide discussion of techniques in neural rendering. Singer et al. \cite{singer2022make} and Ho et al. \cite{ho2022imagen} have proposed methods to generate videos from text descriptions, however, the outputs of these methods are not yet of high enough quality or resolution to be used for generating specific sky scenarios or photorealistic lighting.

\section{Multi-timescale Sky Appearance Prediction}
\label{sec:method}

Static cloud synthesis on the hemisphere for use in rendering can be modeled by existing techniques such as Satilmis et al. \cite{satilmis2022deep} and Mirbauer et al. \cite{mirbauer2022skygan}. However in order to generate realistic cloud movement over multiple frames, we can start with these approaches or captured environment illumination and generate a sequence of images which contain cloud movement. As this needs to cover a longer timescale than a single frame, we need a method which provides smooth cloud movement frame-to-frame but coherent and plausible larger-scale changes to cloud shape, position, and illumination over a longer time period.

\begin{figure}[tp]
    \centering
    \includegraphics[scale=0.5]{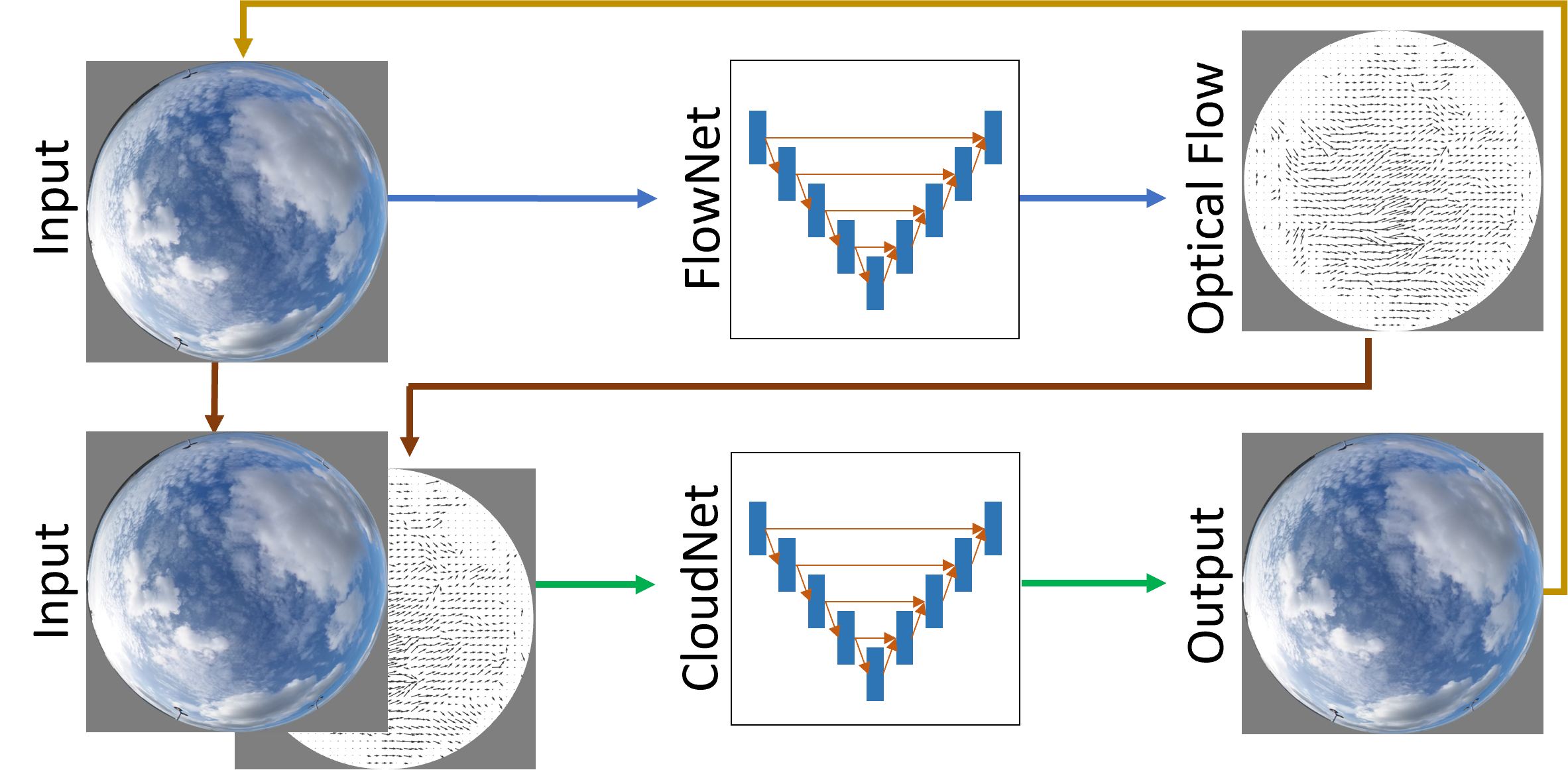}
    \caption{The medium timescale Non-Linear model is a recursive method. $CloudNet$ takes two inputs: a sky image and its optical flow image. $FlowNet$ is used to estimate the optical flow from the sky image. The output is the cloud image after $\Delta t$ times and feedbacked to $CloudNet$ to generate the next output. These recursive predictions create dynamic clouds.}
    \label{fig:overviewCNN}
\end{figure}

Motivated by this, we divide the timescales of cloud evolution into three: long, medium and short term. Long term here refers to the timescale in which the clouds or weather may change substantially and is typically modeled via weather forecasting. Medium term in the context of our work refers to tens of seconds, where the position and shape of the clouds may change substantially, but the types of clouds present are constant. Short term refers to the movement of clouds over a short timescale, in this work, this is less than 10 seconds, as discussed in Section \ref{subsubsec:flowfieldcreation}.

As the focus of this work is on dynamic clouds, we focus on the short and medium timescales to capture movement. This faces the challenge that changes to cloud shape and position have to be smooth and coherent over hundreds of frames, while simultaneously predicting correct illumination. We also have to predict clouds moving at different speeds, due to factors such as different wind strengths for clouds at different altitudes  \cite{hadrich2020stormscapes}. While cloud evolution at these timescales can be predicted by physics-based models such as simulating fluid dynamics in the atmosphere, we operate in image space as this removes the need for complicated and expensive cloud, terrain, and atmospheric models.

However, simply predicting a subsequent frame of sky appearance given a previous image faces the issue that cloud appearance between frames which may be a fraction of a second apart, may change an extremely small amount, but over time the appearance change can be substantial. Therefore, a multi-timescale model is required. We take an approach which predicts the appearance of the sky at the medium scale, and uses a linear model to predict the short term movement conditioned on the current and next medium term prediction.

Specifically, the dynamic sky-lighting, represented as an image $\mathzapf{I^{t}}$, is modeled as a function of time $t\in\mathbb{R}^+$. Input and outputs are represented as an image $I^{t}$ at a given time. The medium term, non-linear part $\xi(I^{i\Delta t})$ predicts the lighting at each $\Delta t$ time intervals and short-term, linear part, $\Gamma(I^{i\Delta t}, I^{(i + 1)\Delta t}, t')$ predicts cloudy sky illumination between each medium term output at time $t' \in (0, .., \Delta t)$. This can be summarized as:

\begin{equation}
    I^{t} =
    \begin{cases}
         \xi(I^{i\Delta t}), &  \quad t \mod{\Delta t} = 0 \\
         \Gamma(I^{i\Delta t}, I^{I(i + 1)\Delta t}, t'), & \quad otherwise \\
    \end{cases}
    \label{eq:overview}
\end{equation}

\noindent where $i = \lfloor t / \Delta t \rfloor$. The input, $I^{0}$, can be any hemispherical image of the sky, either captured or synthesized by a generative approach Satilmis et al. \cite{satilmis2022deep}, Mirbauer et al. \cite{mirbauer2022skygan}. Figure \ref{fig:overview}, provides a summary of the method, and the functions $\xi(I^{t})$ and $\Gamma(I^{t}, I^{t + \Delta t}, t')$ are explained in the following sections.

To achieve this in image space we need an invertible and low distortion mapping from the (hemi)sphere to image space $M: \mathbb{S}^2 \mapsto \mathbb{R}^2$. We use a fisheye mapping as used in previous work by Satilmis et al. \cite{satilmis2022deep}, although our method can be re-trained to work with other mappings. We also need to take into account the dynamic range of real skies when applying our method. Similar to Chen et al. \cite{chen2022text2light}, we map the input High Dynamic Range image into a $[0..1]$ domain, and apply an inverse tone mapper e.g. Marnerides et al. \cite{marnerides2018expandnet}, Khan et al. \cite{cloudimto} to boost the final frame back to the appropriate dynamic range.

\subsection{Medium Timescale Non-Linear Model} \label{subsec:NonLinearModel}

\begin{figure*}[!htp]
    \centering
    \includegraphics[scale = 0.5]{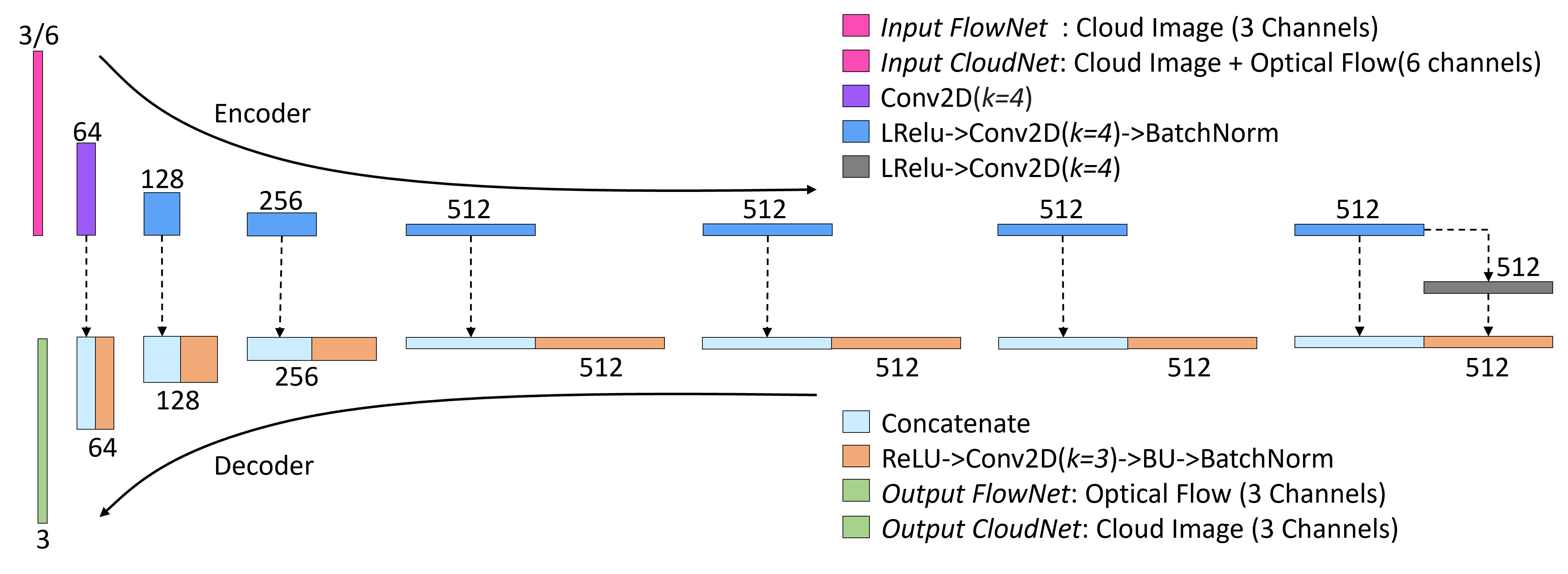}
    \caption{The UNet architecture used in the framework. $CloudNet$ and the $FlowNet$ use the same network structure except for the input layer. $FlowNet$ takes a 3-channel input (cloud image), and $CloudNet$ has a 6-channel input (concatenated cloud image and flow field). The encoder layers encode the inputs by applying eight sequential 4x4 convolutions, then the last seven convolutions come after a Leaky ReLU activation function and the six intermediate ones are followed by batch normalization. The decoder layer decodes the inputs by applying eight sequential ReLU, 3x3 convolutions, bilinear upsampling and batch normalization. All the convolutions use padding and stride of 1.}
    \label{fig:UNet}
\end{figure*}

The medium timescale non-linear approach is composed of two Convolutional Neural Network (CNN) models, see Figure \ref{fig:overviewCNN}. The first model, $FlowNet$ ($\funF$), is trained to predict the movement and shape changes of the clouds at $\Delta t$ intervals. To encode the movement of the clouds in this model, we propose to use a flow field $\in \mathbb{R}^3$ (two channels encode the angle, the other encodes the magnitude) to predict where each cloud pixel will move in the next $\Delta t$ time. The second model, $CloudNet$ ($\funC$), is trained to predict the cloud illumination given a concatenation ($\oplus$) of the previous frame's illumination $I^{t}$ and the flow field predicted by $FlowNet$. This also serves to reconstruct high-frequency details which may have been lost from naively applying a flow field. The overall non-linear model is defined as:

\begin{equation}
    \xi(I^t) = \funC(I^t \oplus \funF(I^t))  
\end{equation}

\subsubsection{CNN Architecture} \label{subsubsec:CNN}

The models used in the framework are both UNets \cite{ronneberger2015u}. This model is chosen for its efficiency in recovering the low-level encoded data and also its success in generating cloud images \cite{satilmis2022deep}. 

$FlowNet$ and $CloudNet$ share the same network structure except for the input layer, see Figure \ref{fig:UNet}. $FlowNet$ takes a cloud image with 3 channels as input at $I^{i \Delta t}$ and outputs a predicted flow field $\vec{I}^{i \Delta t}$. $CloudNet$ takes two inputs: the same cloud image at the previous timestep, $I^{i \Delta t}$ and the flow field $\vec{I}^{i \Delta t}$ resulting from $FlowNet$, which are concatenated to 6 channels. The encoder uses 2D convolutional layers with $kernel size=4$, $stride=1$ and $padding=1$ to downsample the images. The decoder uses bilinear upsampling, which doesn't suffer from checkerboard effects that might appear in the output, see Marnerides et al. \cite{marnerides2018expandnet}. This is followed by 2D convolutional layers with $k=3, s=1, p=1$. Following a similar structure to Isola et al. \cite{isola2017image}, LRelu with slope $0.2$ and ReLU are used as activation functions in the encoder and decoder, respectively. The networks have 64-128-256-512-512-512-512 features sequentially.

\begin{figure*}[!htp]
    \centering
    \includegraphics[scale=0.5]{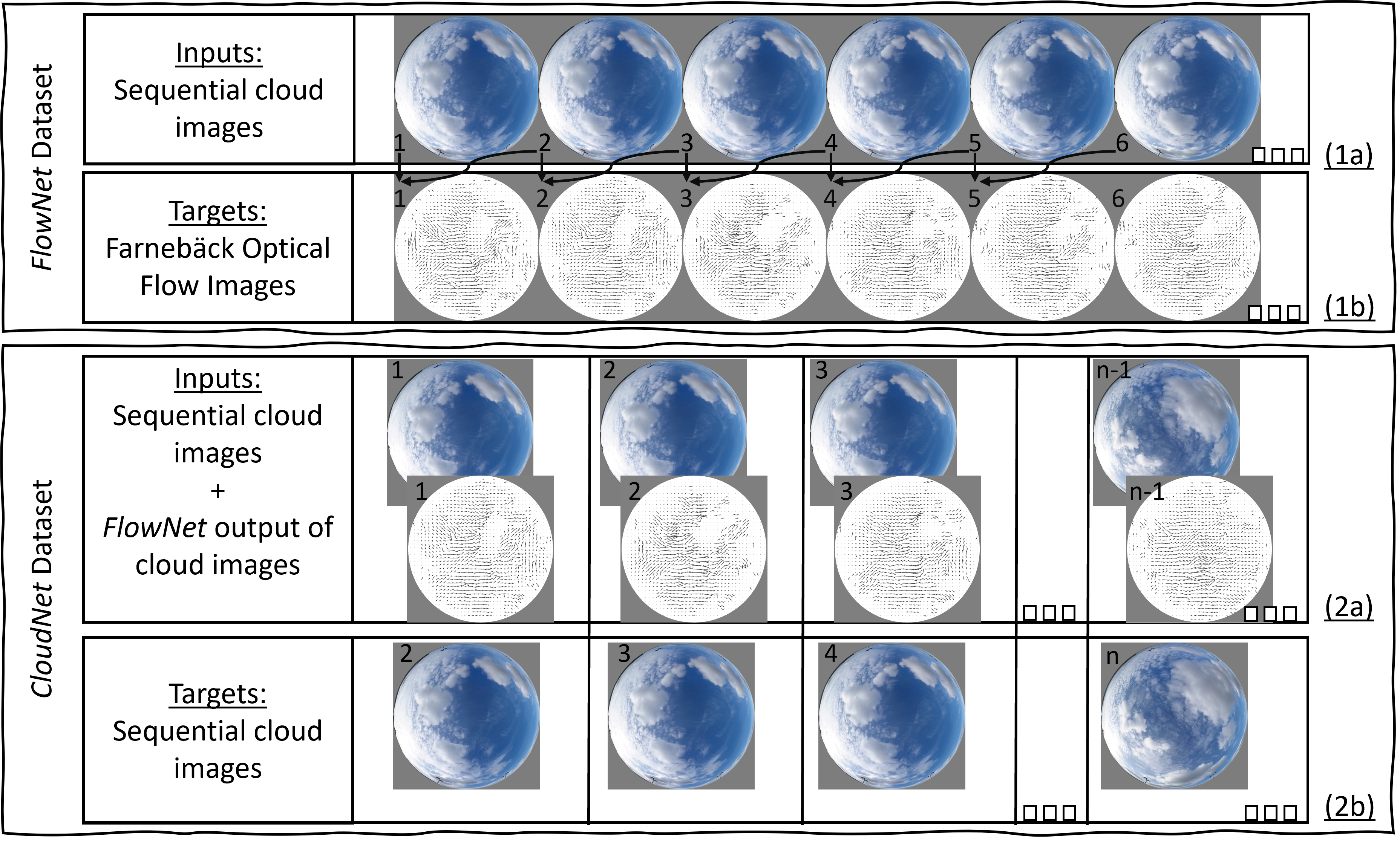}
    \caption{Dataset for $FlowNet$ and $CloudNet$. Our base dataset is composed of whole sky images captured at 10-second intervals (1a). For each image, a flow field is created using the method by Farneb{\"a}ck et al. \cite{farneback2003two} (1b) to estimate the optical flow between each image and the consecutive image. Each sky image and its optical flow estimation forms the $FlowNet$ dataset. The $CloudNet$ input dataset is composed of the base pixel data and the corresponding output from $FlowNet$ (2a). The target images are consecutive input sky images (2b).}
    \label{fig:dataset}
\end{figure*}

\subsubsection{Flow Field Creation}
\label{subsubsec:flowfieldcreation}

The flow fields, $\vec{I}^{i \Delta t}$, in this work are generated by using optical flow, specifically the approach by Farneb{\"a}ck et al. \cite{farneback2003two}. This produces an initial flow field $\vec{I}^{i \Delta t}_{F}$. As initial investigations showed that the optical flow method failed to estimate the flow if the time interval is too large between the images, we chose $\Delta t = 10$ seconds, which we found balanced between capturing enough movement and allowing the optical flow algorithm to produce valid results.

To ensure that optical flow is applied only to the cloud pixels, we estimate a binary cloud mask $I^{i\Delta t}_{cm}$ and elementwise multiply ($\odot$) the initial flow field by this mask to produce the final flow field:

\begin{equation}
    \vec{I}^{i \Delta t} = \vec{I}^{i \Delta t}_{F} \odot I_{cm}
\end{equation}

The cloud mask $I_{cm}$ is computed for each pixel by:

\begin{equation}
    I^{i\Delta t}_{cm} = 
    \begin{cases}
        1 & if \frac{I^{i\Delta t}(R)}{I^{i\Delta t}(B)} > 0.46 \\
        0 & otherwise,
    \end{cases}
    \label{eq:Icm}
\end{equation}

\noindent where $I^{i\Delta t}(R)$ and $I^{i\Delta t}(B)$ are the red and blue color channels of $I^{i\Delta t}$ respectively. The thresholded ratio of these color channels $\frac{I^{i\Delta t}(R)}{I^{i\Delta t}(B)}$ is used for identifying pixels consisting of clouds. This is a common method used in cloud-sky segmentation \cite{johnson1987automated}. We found the value of $0.46$ worked well for thresholding cloud pixels from clear sky pixels in all of our dataset.

\subsubsection{Dataset and Training} \label{subsubsec:dataset_training}

% \begin{figure}[!h]
%     \centering
%     \includegraphics[scale=0.4]{Images/DatasetExamples.png}
%     \caption{Some cloud image examples from the dataset}
%     \label{fig:dataset}
% \end{figure}

The dataset is composed of 3626 sequential LDR images captured every $\Delta t$ seconds. 80$\%$ of the images were used in training and the rest were used for testing. The resolution of the images is 2048x2048. Images were captured in the UK with a Ricoh Theta Z1 \cite{RicohThetaZ1}, see Figure \ref{fig:dataset}. A flow field for each image was computed via the method outlined in Section \ref{subsubsec:flowfieldcreation} and used for training $FlowNet$.

During the training phase of $CloudNet$ and $FlowNet$, all the input sky-images are taken from the dataset ($t=0$). The optical flow inputs used for $CloudNet$ training are generated by using the trained $FlowNet$ model, see Figure \ref{fig:dataset} for an illustration of this process.

\subsubsection{Loss Function} \label{subsubsec:LossFunc}

For the loss function, the mean squared error (MSE) and cosine similarity loss functions were used. MSE is chosen as it is good at recovering the encoded data, but can lead to images which lack fine details. Furthermore, it provides low prediction accuracy in color distribution of pixel values \cite{satilmis2022deep}. Therefore, similar to Satilmis et al. \cite{satilmis2022deep}, we included Cosine similarity due to its success in learning RGB values.

\subsection{Short Timescale Linear Model} \label{subsec:LinearModel}

The short timescale model produces smooth cloud movement between every cloud image generated by the medium time non-linear model, i.e. the output of $CloudNet$. To ensure consistent and smooth cloud movement, the short timescale model must match the output of $CloudNet$ at times $i\Delta t$ and $(i + 1)\Delta t$. Furthermore, the flow fields at these times should also match to ensure consistent movement of these clouds. While it would be tempting to formulate this short-timescale model as a differential equation with Cauchy boundary conditions, in this case defined at times $i\Delta t$ and $(i + 1)\Delta t$, the lack of uniqueness of the solution in this context limits practical application to this problem. Therefore, we propose a simple, fast-to-compute, deterministic model which respects the boundary conditions and leads to smooth cloud movement.

To achieve this, we split the short-timescale into three sub-timescales and solve cloud movement via a piecewise linear model. At times $(i + \frac{1}{3})\Delta t$ and $(i + \frac{2}{3})\Delta t$, two intermediate frames are computed by advecting the clouds in image space along the respective flow fields forward in time from $i\Delta t$ and backwards in time from $(i + 1)\Delta t$ and then linearly interpolating between these images in the intermediate times. We define a function $\Lambda(I, \vec{I}, t')$, which advects the cloud pixels from $I$ along a flow field $\vec{I}$ by an amount proportional to the time $t'$ which takes a value between $i\Delta t$ and $(i + 1)\Delta t$. Interpolation is performed by another function $\Upsilon(I^{a}, I^{b}, t')$ which interpolates between two images at times $a$ and $b$ with $t'$ defined as before.

\begin{figure*}[!htp]
    \centering
    \includegraphics[scale=0.31]{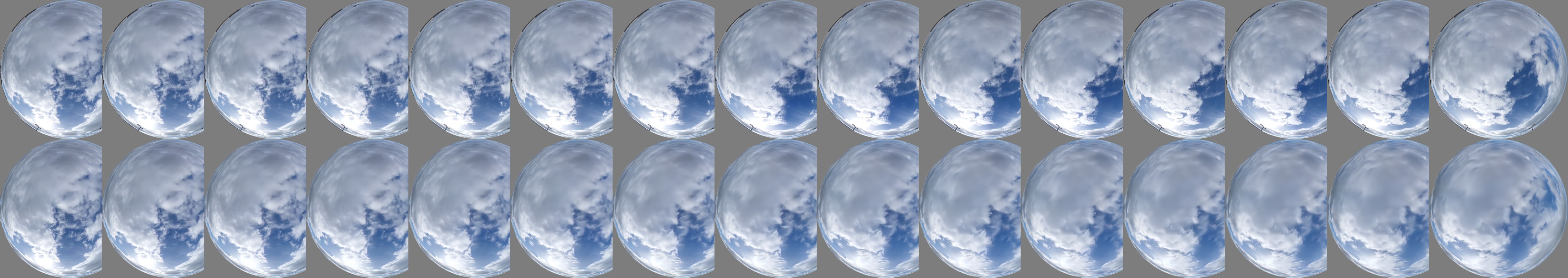}\\ \vspace{1mm}
    \includegraphics[scale=0.31]{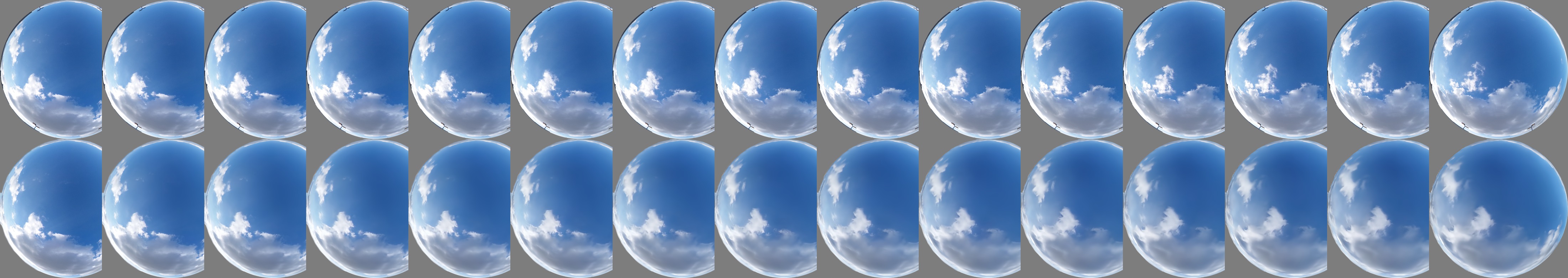}\\ \vspace{1mm}
    \includegraphics[scale=0.31]{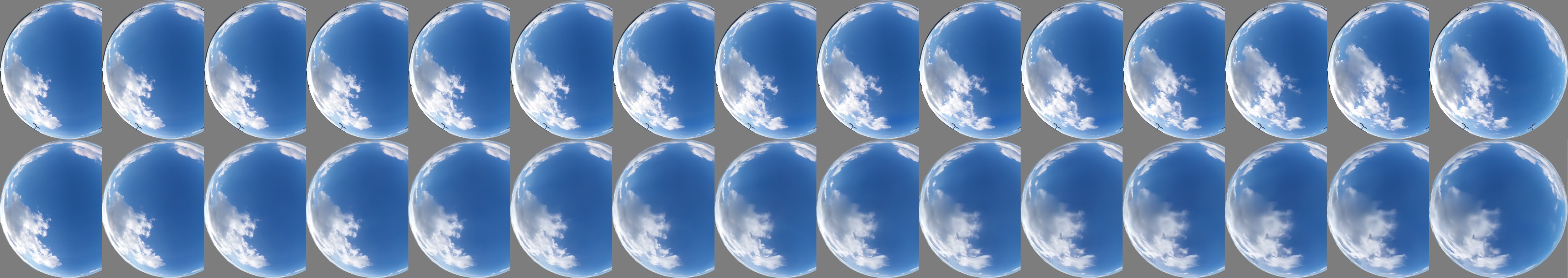}
    \caption{Visual comparison of our method with a ground truth dataset. The first row is the test set and the second row are the predictions from our model. This shows that approaches generate similar predictions in terms of structure and illumination.}
    \label{fig:seqRealGen}
\end{figure*}

The motivation behind this is to ensure the boundary conditions are respected which is automatically the case based on advecting the clouds along the flow field forwards and backwards, and we found that the interpolation in the middle step was sufficient to blend the outputs thus avoiding any discontinuity in cloud movement or appearance. This can be summarised as follows:

\begin{equation}
\begin{aligned}
    \Gamma(I^{i \Delta t}, I^{(i+1)\Delta t}, t') = \hspace{5cm}\\
    \begin{cases}
        \Upsilon(I^{i \Delta t},I^{(i+\frac{1}{3}) \Delta t},t'), &  t' \in (0, \frac{\Delta t}{3}) \\
        \Lambda(I^{i \Delta t}, \vec{I}^{i \Delta t}, \frac{\Delta t}{3}), &  t'= \frac{\Delta t}{3}  \\
        \Upsilon(I^{(i+\frac{1}{3}) \Delta t},I^{(i+\frac{2}{3}) \Delta t},t'), &  t' \in (\frac{\Delta t}{3}, \frac{2\Delta t}{3}) \\
        \Lambda(I^{(i + 1) \Delta t}, -\vec{I}^{i \Delta t},\frac{\Delta t}{3}), &  t' = \frac{2\Delta t}{3}   \\
        \Upsilon(I^{(i+\frac{2}{3}) \Delta t},I^{(i+1) \Delta t},t'), &  t' \in (\frac{2\Delta t}{3}, \Delta t) \\
    \end{cases}
\end{aligned}
\end{equation}

\noindent where $I^{i \Delta t}$ and $I^{(i+1)\Delta t}$ are two sequential predictions of the $CloudNet$ and $\vec{I}^{i \Delta t}$ is the optical flow between $I^{i \Delta t}$ and $I^{(i+1) \Delta t}$.

Moving clouds linearly along these vectors in image space does not take into account the curvature of the sphere. Therefore, the vectors in image space have to be converted to the sphere via the inverse mapping $M^{-1}$. Interpolation is performed along a great arc on the sphere, then mapped back to image space. In practice, we re-estimate the optical flow at timesteps $i\Delta t$ and $(i + 1)\Delta t$ to incorporate the extra high frequency details which are added by $CloudNet$ and use the method discussed in Section \ref{subsubsec:flowfieldcreation} to ensure that homogenous regions within the large clouds are filled with plausible values.
\section{Results} \label{Sec:Results}

In this section, we present the results of our method. First, we show qualitative and quantitative results for the dataset described in Section \ref{subsubsec:dataset_training}. We compare against real-world data, as to the best of our knowledge, there is no similar approach for generating whole sky dynamic clouds in the literature. Then, we provide examples of temporal clouds generated with synthesized inputs created with an existing generative cloud model. Lastly, we demonstrate results showing the main use case of our method which is rendering 3D scenes with generated cloudy sky animations. Further results generated with our method can be seen in the video in the supplementary material.

\subsection{Real-world results} \label{subsec:results_nonlinear}

The accuracy of the $FlowNet$ and $CloudNet$ is evaluated through the test set of $20\%$ of the dataset (725 frames). The structure of this set is described in Section \ref{subsubsec:dataset_training}. We present objective metrics averaged over the sequences in the test set. We show three metrics: MSE and PSNR measure pixel differences of the reconstruction, and SSIM measures differences in structure. In this evaluation for each test image, its next image is compared with $CloudNet$ output. This shows the success of the model in prediction with an error of less than 0.5$\%$. Table \ref{table:results} summarises these results.

\begin{figure}[!tp]
    \centering
    \includegraphics[scale=0.5]{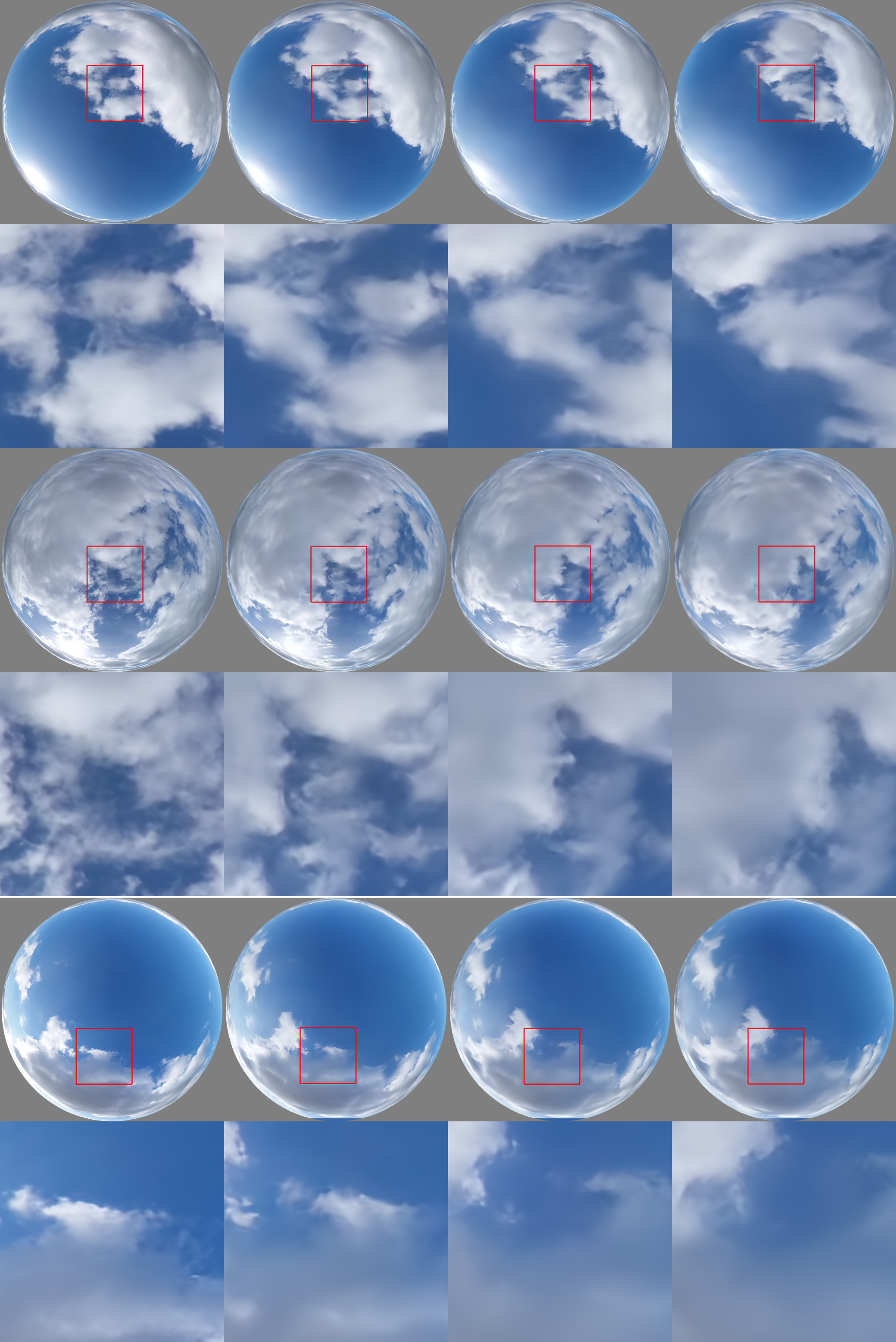}
    \caption{Results of our method shows at $40$ second intervals. This shows our method generates plausible cloud structure and illumination with time.}
    \label{fig:seqzoomed}
\end{figure}

\begin{table}[!h]
\centering
\begin{tabular}{|l|l|l|l|l|}
\hline
      & MSE     & PSNR & SSIM  \\ \hline  % & Cos   
Error & 0.00536  & 23.673  &  0.904   \\ \hline  % & 1.0812 
\end{tabular} \vspace{1mm}
\caption{Summary of performance of our method averaged over the test set of sequences.}
\label{table:results}
\end{table}

We also visually compare our predictions with the ground truth captures in Figure \ref{fig:seqRealGen}. The top row shows ground truth captures, and the bottom row shows our method. This shows our method can generate plausible evolution of the clouds with time and broadly similar illumination generated with $CloudNet$.

\begin{figure}[!tp]
    \centering
    \includegraphics[scale=0.4]{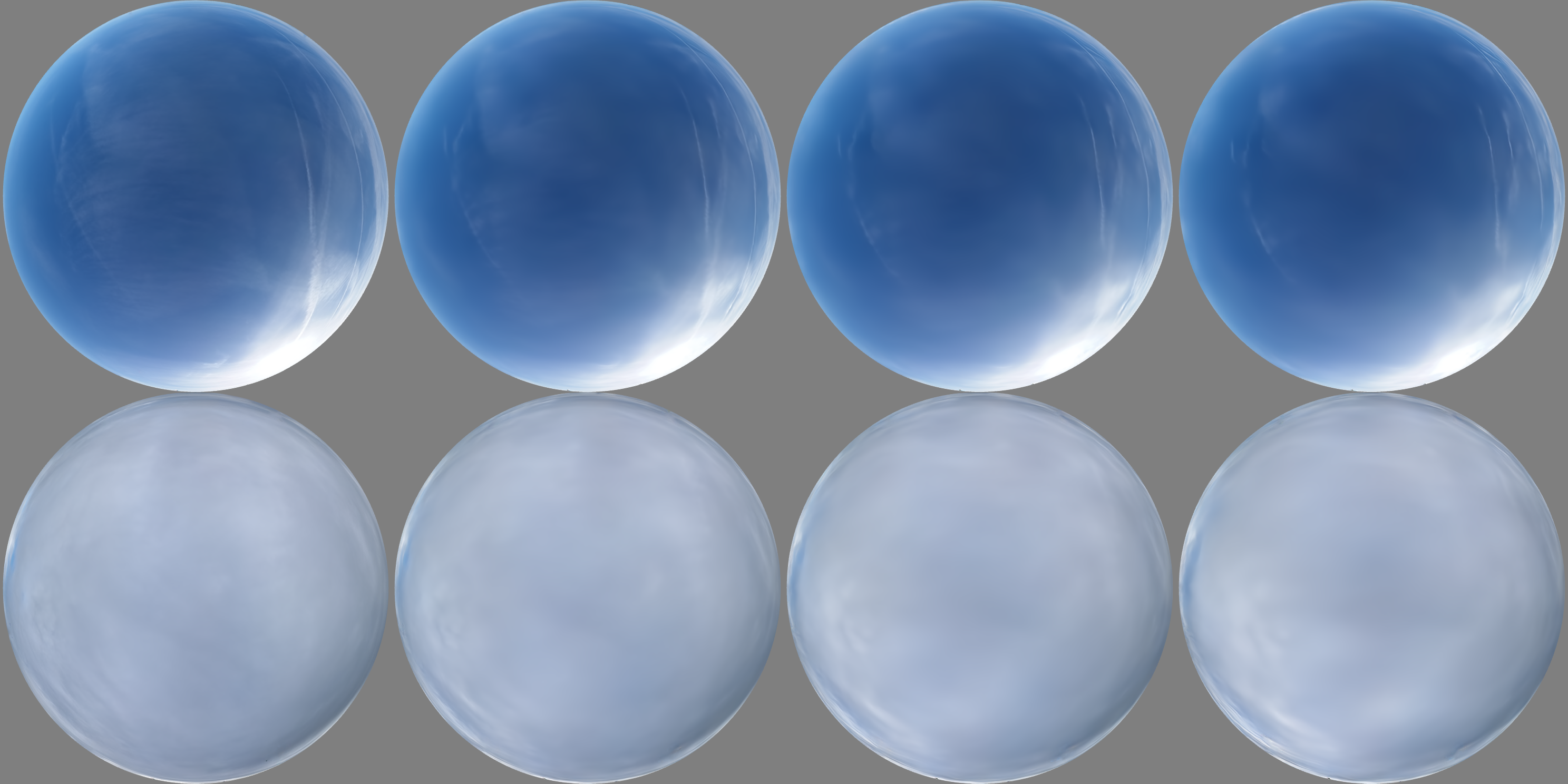}
    \caption{Results showing out method applied to cloudy skies at two extremes. The top row shows our method applied to barely visible cirrus clouds, and at the other extreme the bottom row shows the results of our approach on overcast skies.}
    \label{fig:cirrusovercast}
\end{figure}

Figure \ref{fig:seqzoomed} illustrates predictions of the non-linear model at every $40$ seconds. The rectangular areas show a zoomed view of the generated clouds. This illustrates several features of our model. The clouds are moving in time across the sky and changing their structure, and the generated clouds are illuminated in accordance with the sky and sun position. We also show results where our method can be used to predict barely visible light clouds, such as cirrus with low coverage, or overcast skies at the other extreme. Examples of these types of skies generated with our method can be seen in Figure \ref{fig:cirrusovercast}.

\begin{figure}[!tp]
    \centering
    \includegraphics[scale=0.7]{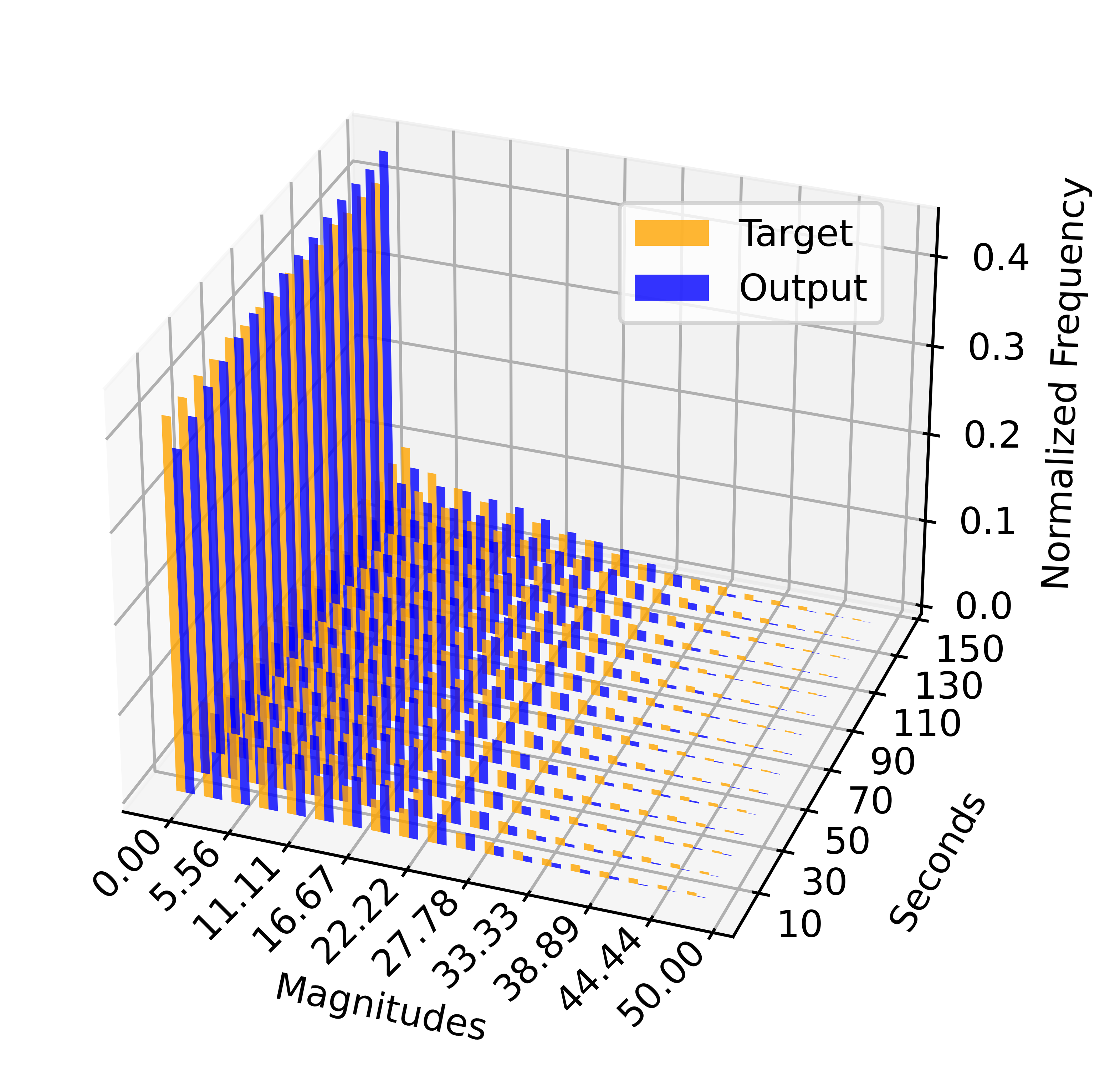}
    \caption{The normalized distribution of the magnitude of the flow field in image space for each medium-term non-linear prediction per frame. These magnitudes correspond to angles on the sphere via the mapping $M^{-1}$. This compares the Target real images (orange) and the Output of our model (blue) and is calculated over all the test images in dataset. This shows that our approach follows a similar distribution to real-world cloud flow, and most of the errors of our approach are concentrated in sub-pixel movements which are less important than large-scale movement.}
    \label{fig:histogram}
\end{figure}

While the previous results assess the visual quality of our results, we also need to evaluate the temporal quality of our method. Our method produces plausible clouds, but the directional movement may not exactly match real data. Minor differences early on in a sequence can lead to exponential differences in cloud positions in later frames, although illumination and shape remain plausible. This is expected behavior, but direct pixel-to-pixel comparisons between frames would lead to meaningless comparisons. However, the magnitudes of the cloud movement frame-to-frame, in our case the magnitude of the flow field, can be expected to be similar between real and synthesized data as this captures how much the clouds move with time.

To assess this, we create a histogram of the magnitudes of cloud movement estimated with optical flow from both the captured real data and the output of our model. Each bin of the histogram stores a range of estimated flow values, from much less than a pixel to multiple pixels\footnote{This also corresponds to movement over a great arc on the hemisphere}, and we plot this with respect to time. This is shown in Figure \ref{fig:histogram} which demonstrates that our model can approximate the amount of flow observed in real clouds as is shown by the similar distribution between real-world flow data (orange) and our predictions (blue). Most of the error between the method corresponds to cloud movements which are much less than a pixel, which typically corresponds to a fraction of a degree on the hemisphere. This shows the success of our model in predicting cloud movement over time.

\begin{figure*}[!tp]
    \centering
    \includegraphics[scale=0.45]{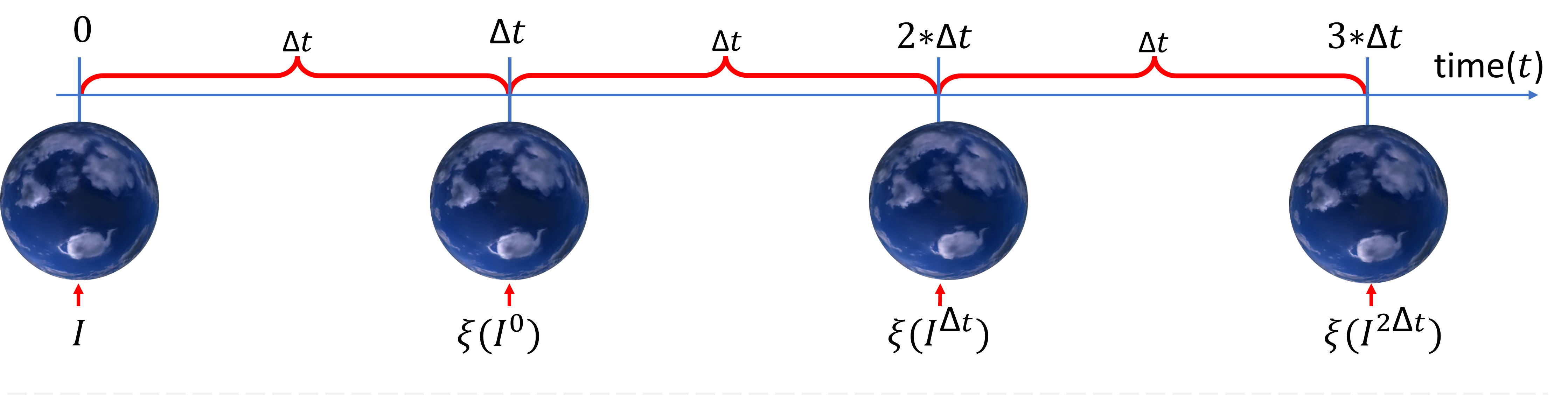}
    \includegraphics[scale=0.45]{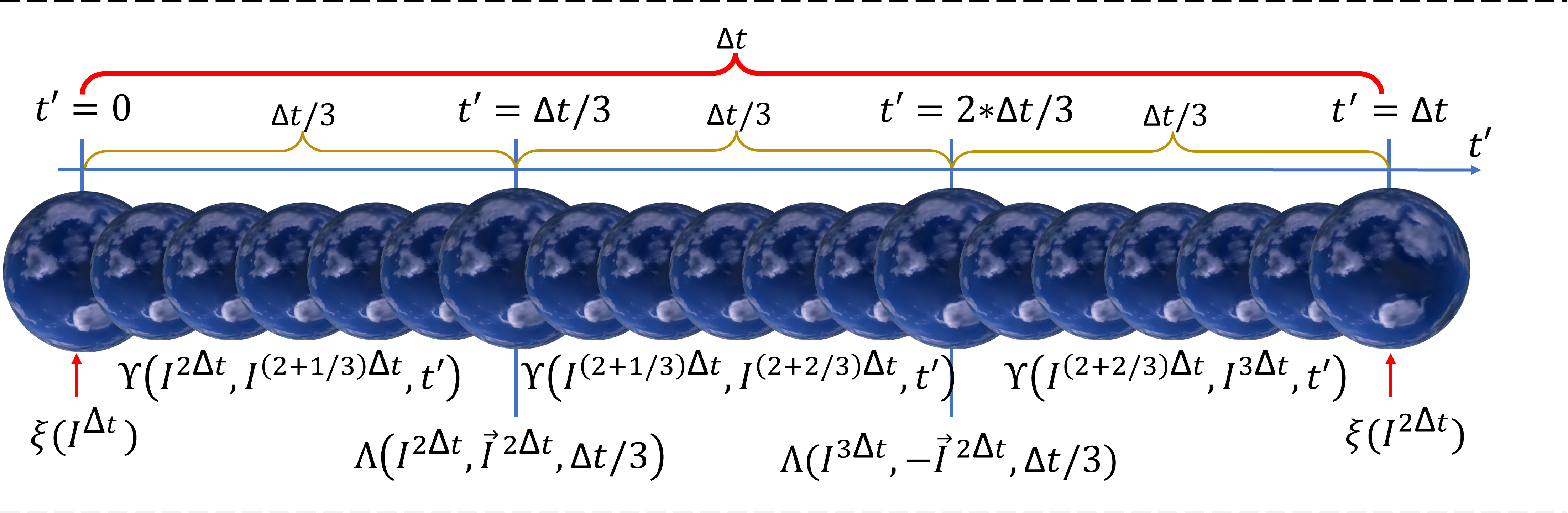}
    \caption{Illustration of results obtained from the combined model. The top row shows the output of the medium term non-linear model which predicts coherent movement over many frames, and the bottom row shows the short term linear model smoothly interpolating between the medium term model outputs.}
    \label{fig:wholemodel}
\end{figure*}

We also show a summary of our method in Figure \ref{fig:wholemodel} which illustrates both the medium term non-linear model and the short-term linear model. This illustrates how our method generates coherent cloud movement over a longer time period via the medium-term model described in Section \ref{subsec:NonLinearModel} and the smooth short-term interpolation approach discussed in Section \ref{subsec:LinearModel}.

\subsection{Deep Synthesized Inputs} \label{subsec:res_deepInputs}

\begin{figure}[!tp]
    \centering
    \includegraphics[scale=0.225]{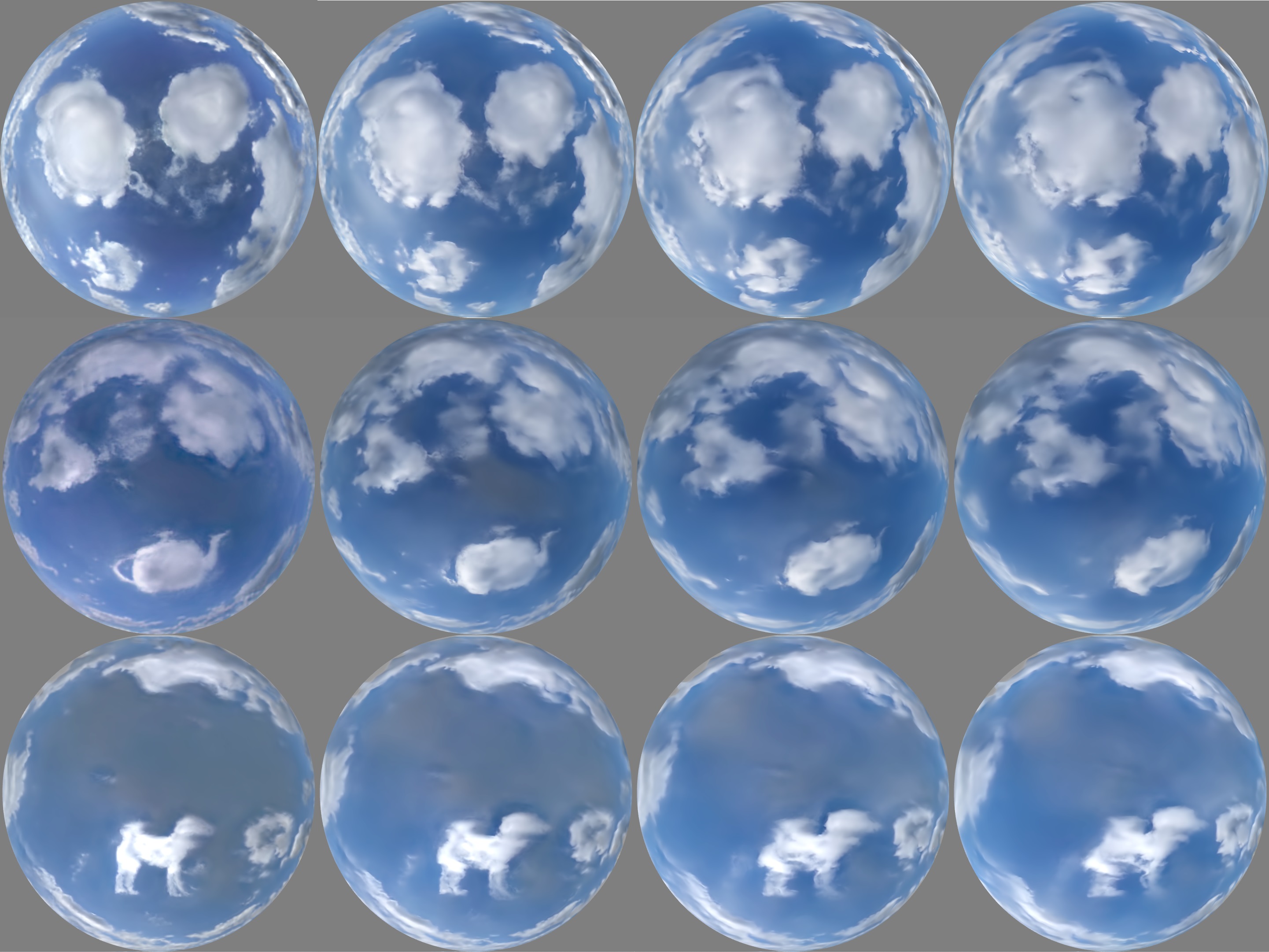}
    \caption{This figure illustrates the prediction of cloud movements by using the inputs generated with the deep synthesis method proposed by Satilmis et al. \citep{satilmis2022deep}. The first images are the inputs, and others illustrate the movement of the clouds after 50 seconds.}
    \label{fig:prevgenerated}
\end{figure}

Our model can also be used to predict cloud movement from static images synthesized by previous methods of generating clouds. This uses the output of these methods as the first frame and then generates cloud movement over time. 

We show results from the method proposed by Satilmis et al. \citep{satilmis2022deep} as their method allowed the use of masks for generating clouds, and thus we can view the evolution of artistically controlled clouds. Figure \ref{fig:prevgenerated} shows three examples of sequences of clouds starting from a generated cloudy sky including clouds in the shape of a ``thought bubble", a teapot, and a dog.

This shows that our method can generate plausible cloud dynamics and illumination, even when the initial cloudy sky was implausible and artist specified.

\subsection{Rendering} \label{subsec:res_rend}

\begin{figure*}[htp]
\begin{tabular}{@{}l@{}l@{}l@{}l@{}l}
\rotatebox[origin=c]{90}{Tower} &
    \includegraphics[scale=0.06666667]{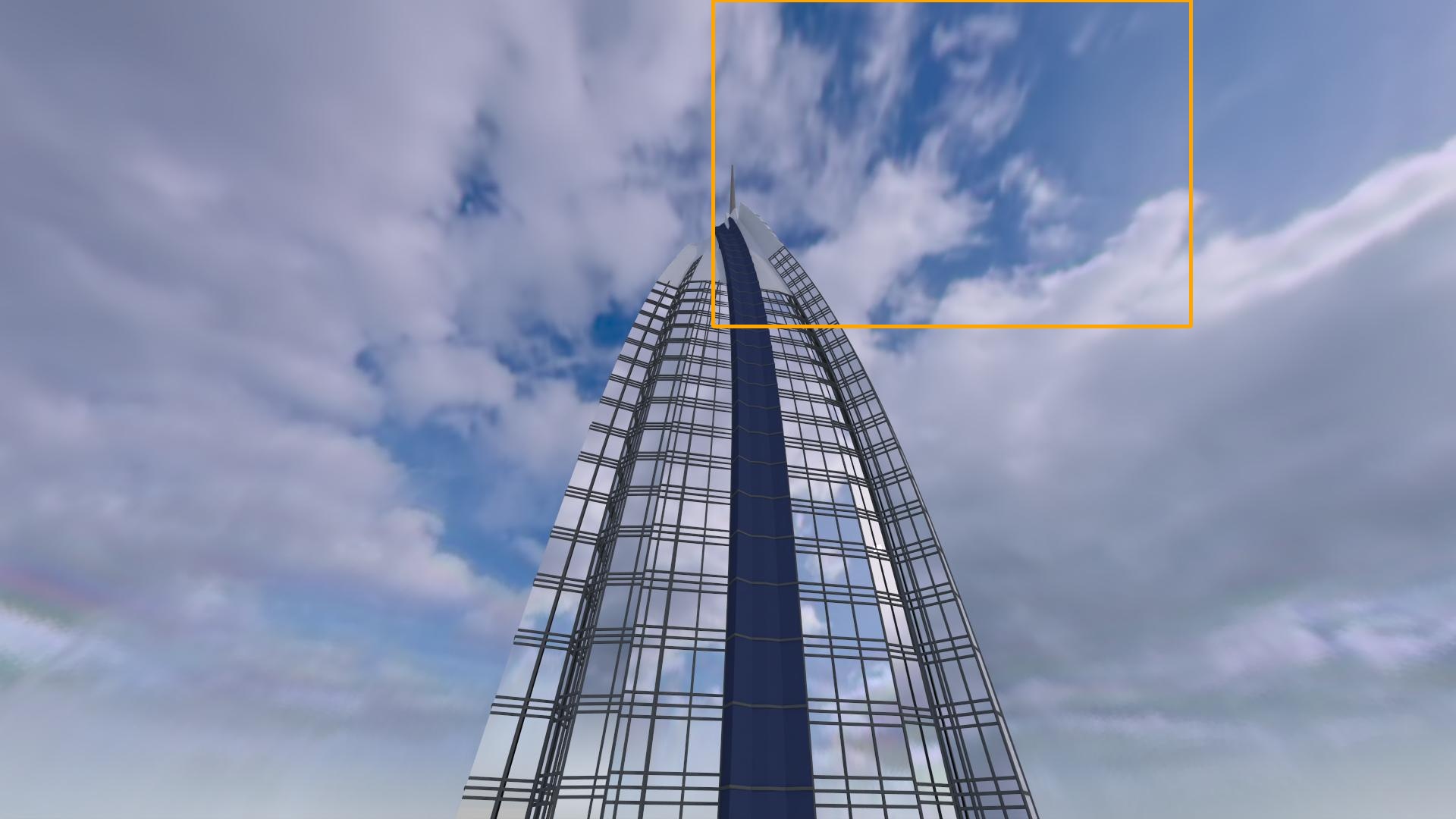} &
    \includegraphics[scale=0.06666667]{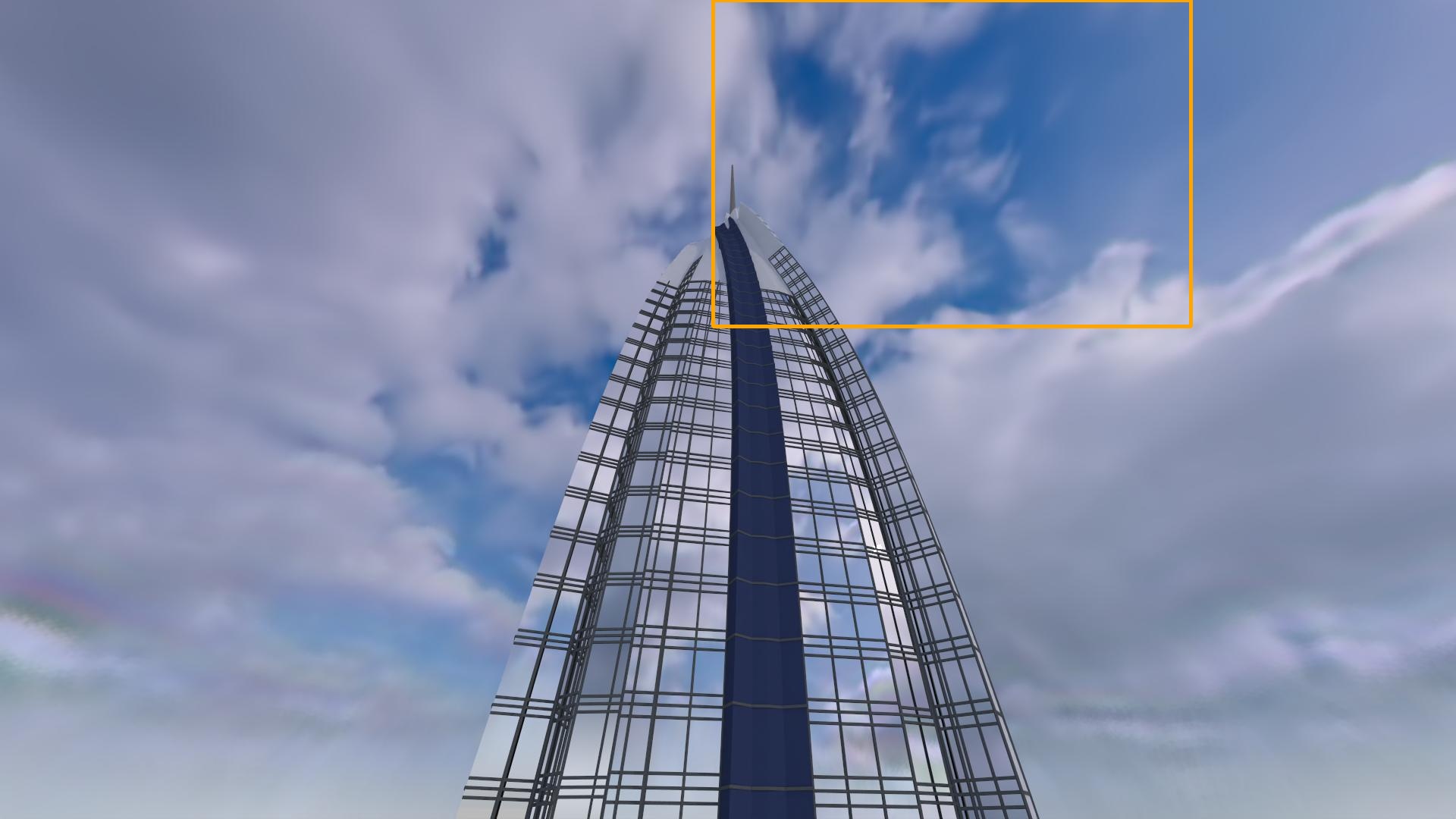} &
    \includegraphics[scale=0.06666667]{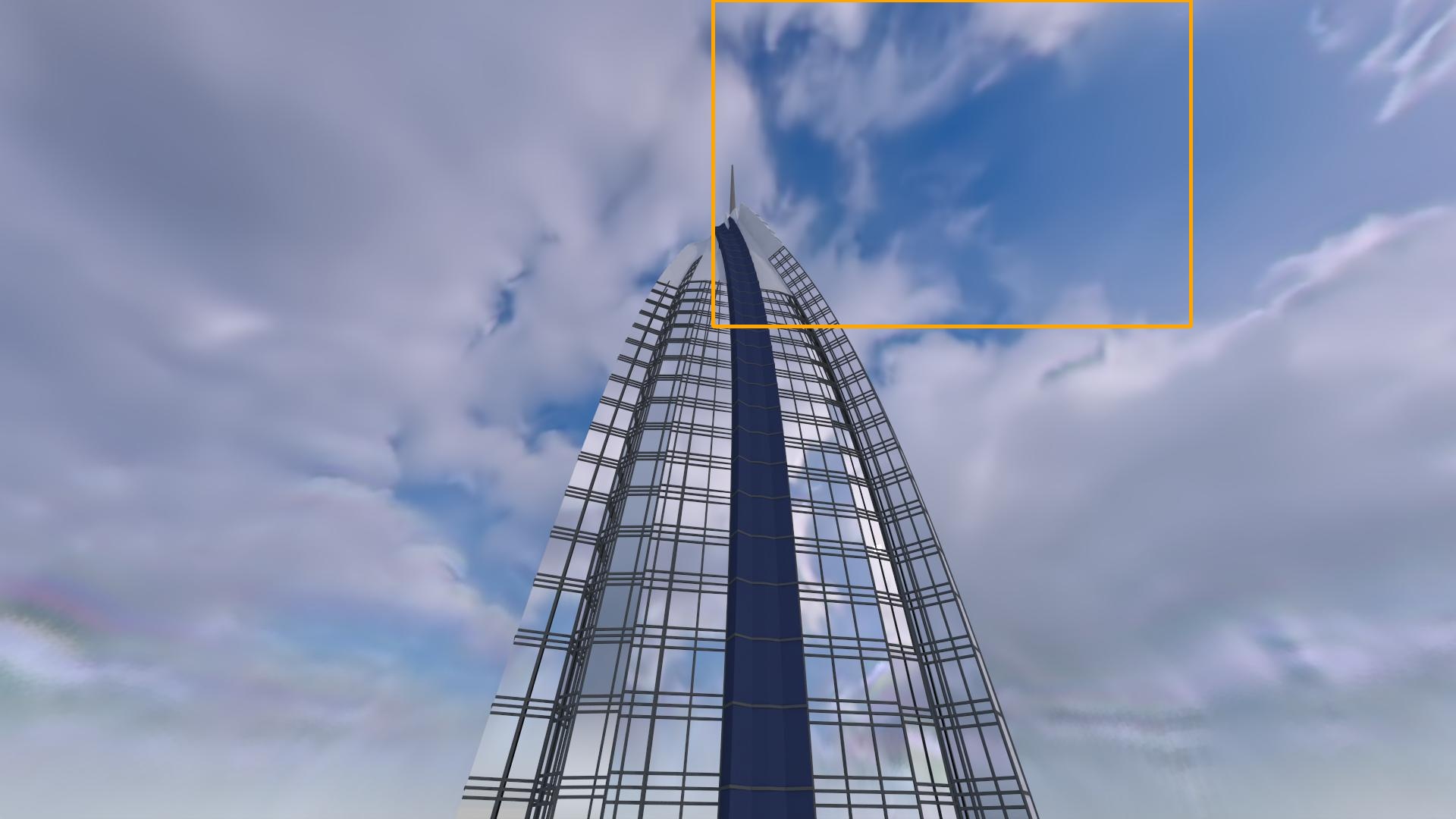} &
    \includegraphics[scale=0.06666667]{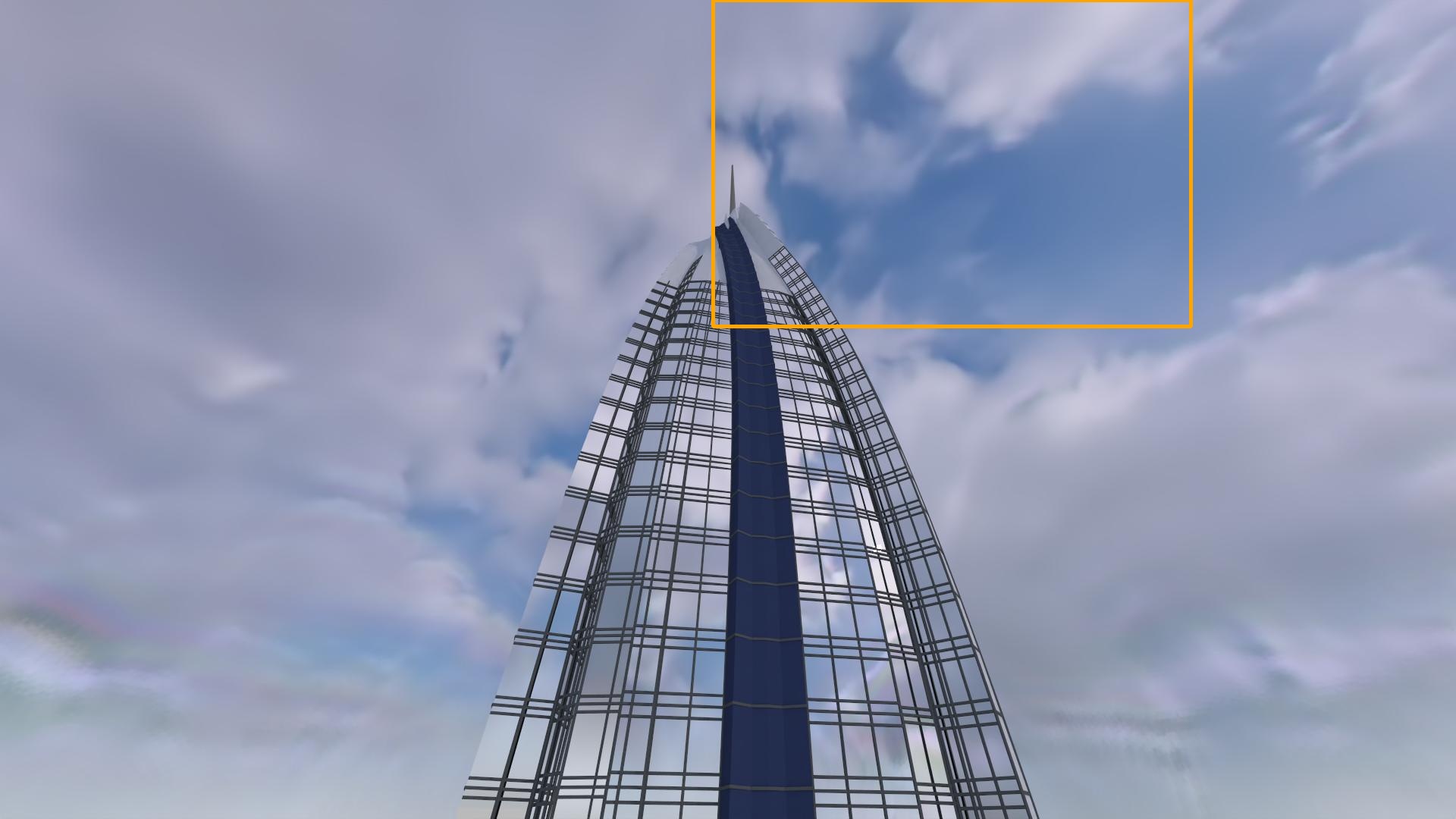} \\ [-3ex]
    & 
    \includegraphics[scale=0.20305]{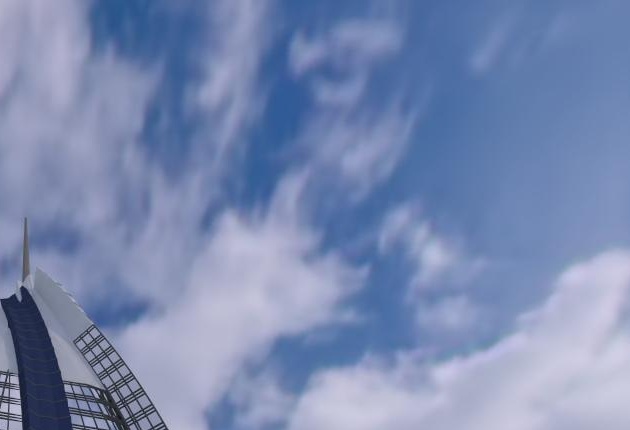} &
    \includegraphics[scale=0.20305]{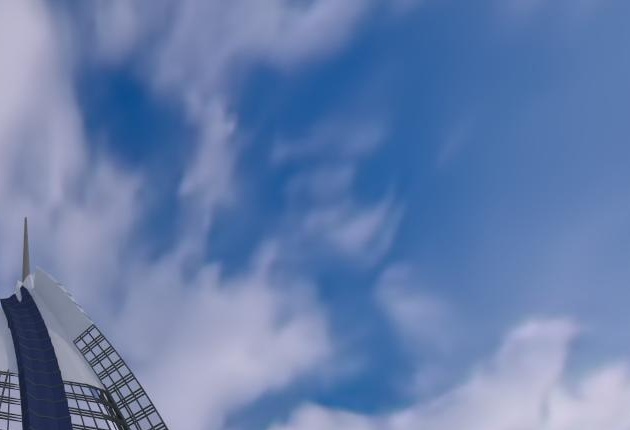} &
    \includegraphics[scale=0.20305]{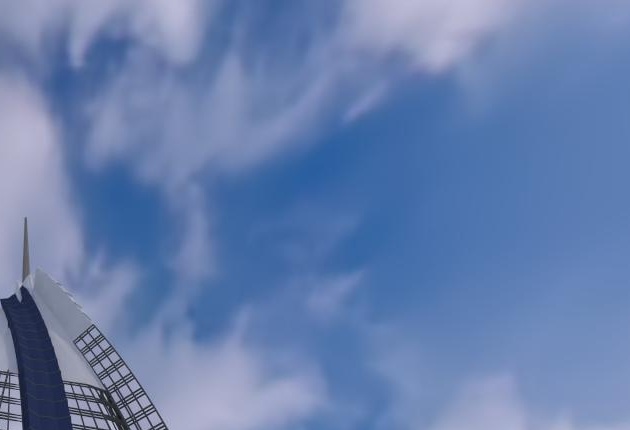} &
    \includegraphics[scale=0.20305]{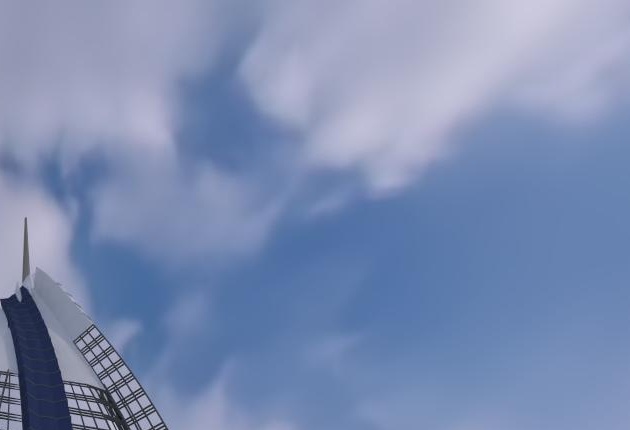} \\
    \rotatebox[origin=c]{90}{House} &
    \includegraphics[scale=0.1]{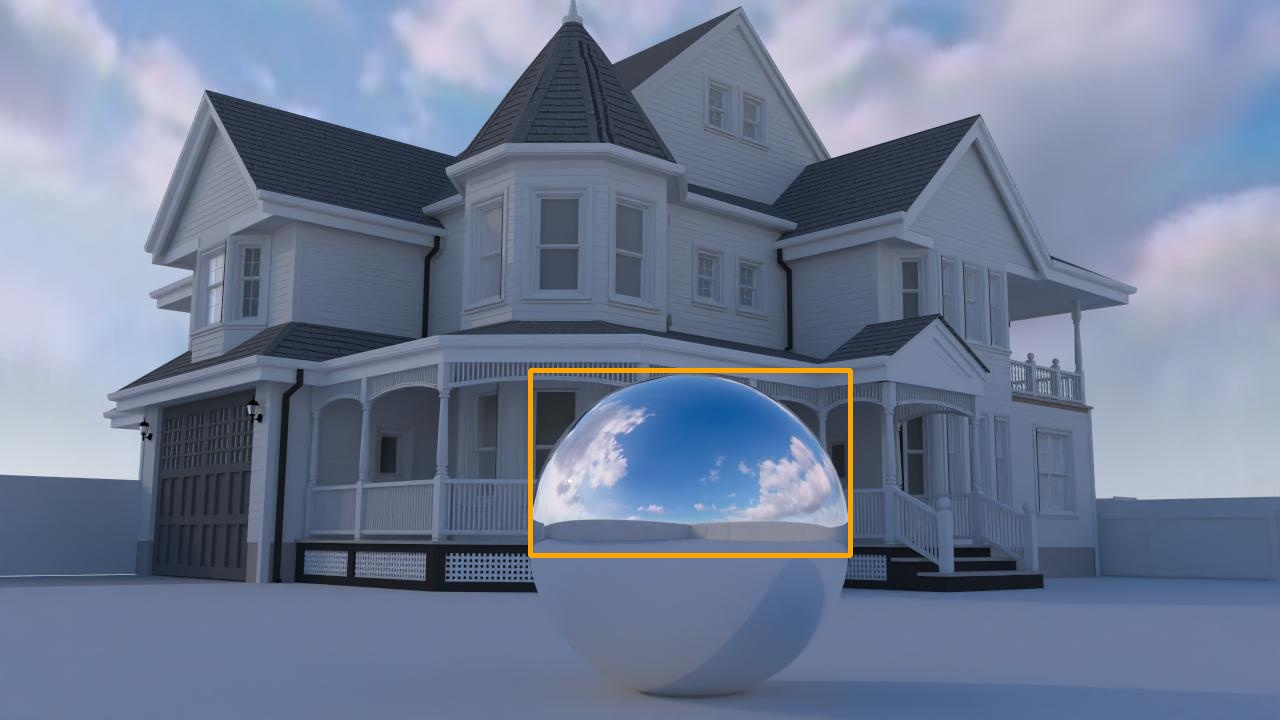} &
    \includegraphics[scale=0.1]{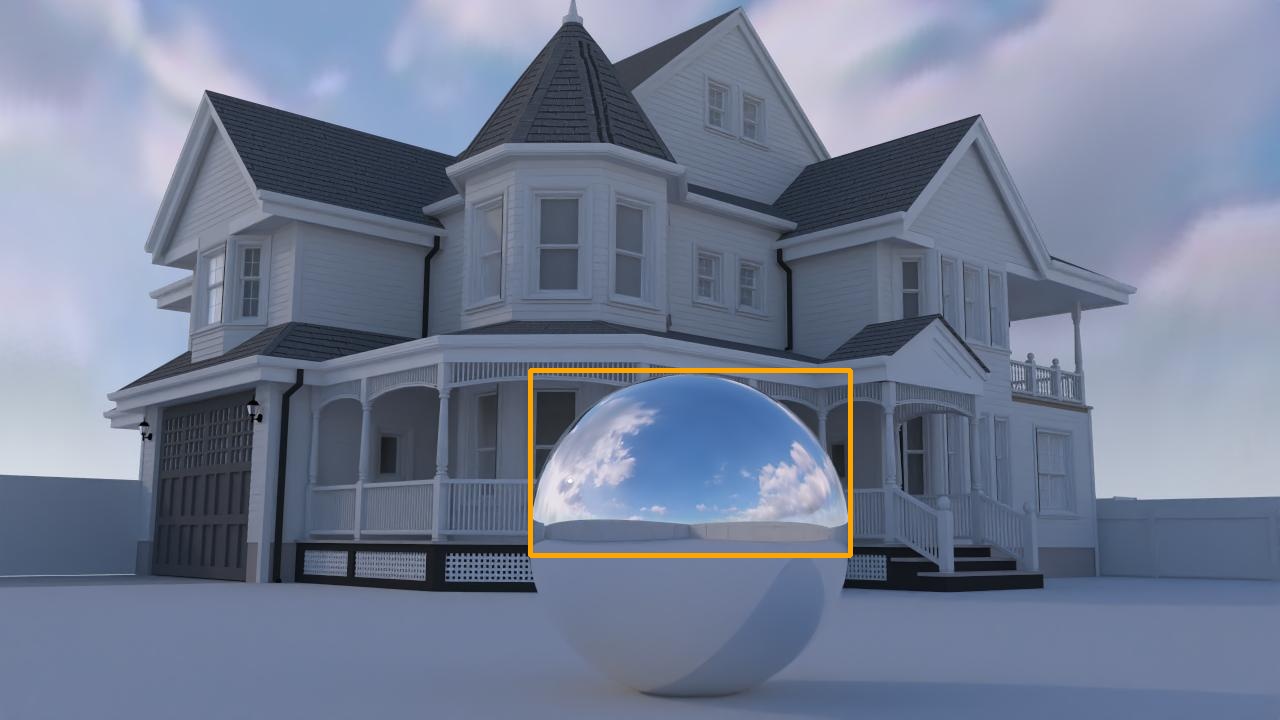} &
    \includegraphics[scale=0.1]{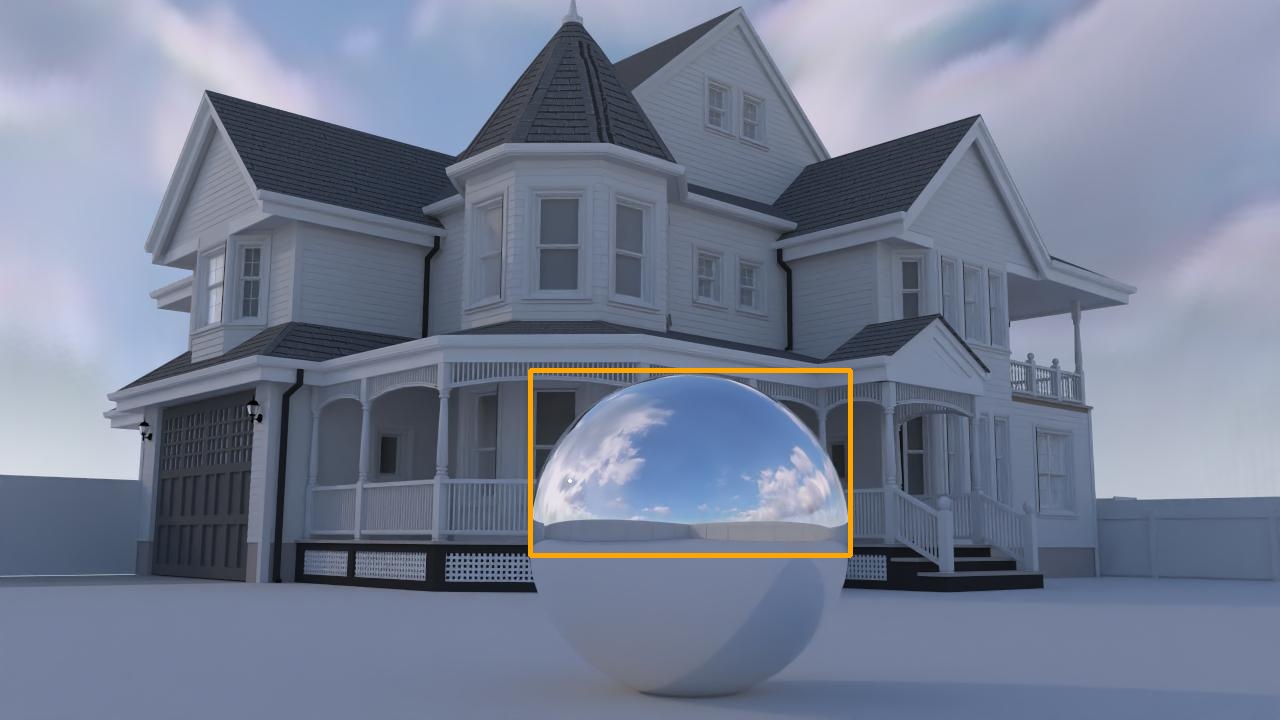} &
    \includegraphics[scale=0.1]{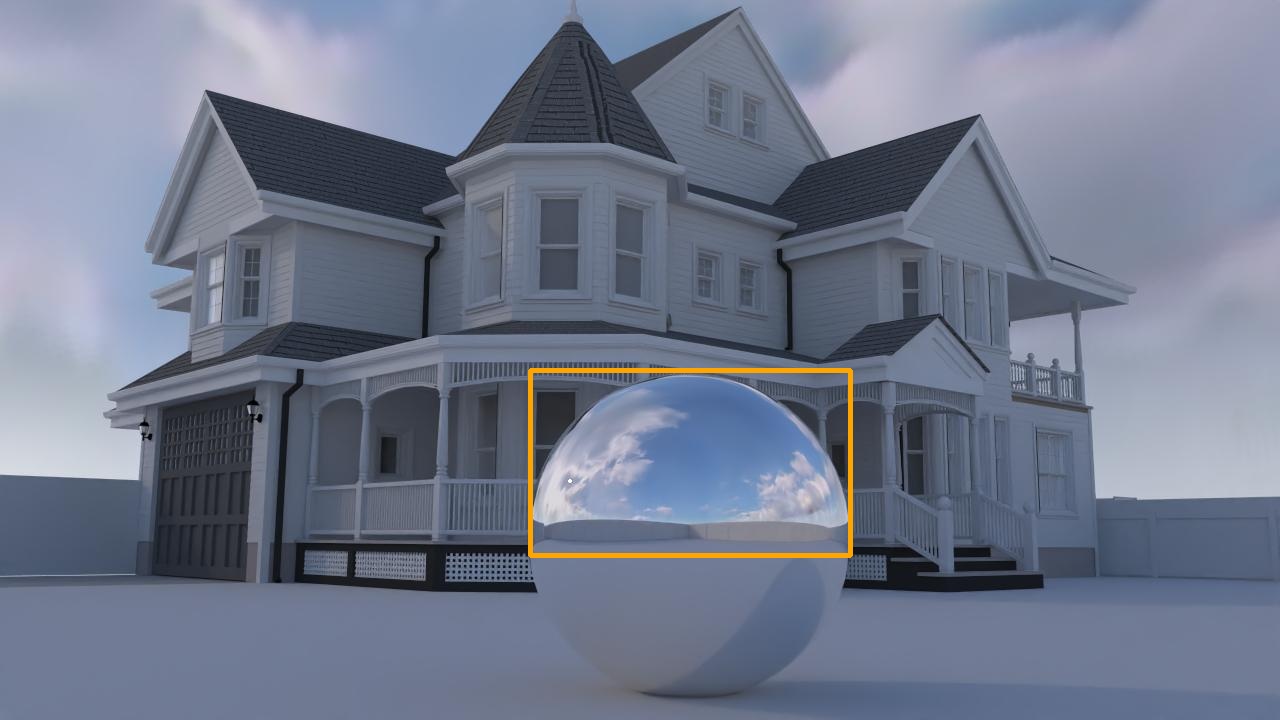} \\ [-5ex]
    & 
    \includegraphics[scale=0.4]{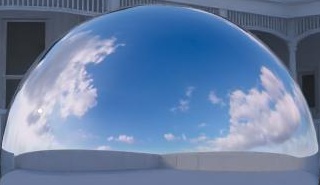} &
    \includegraphics[scale=0.4]{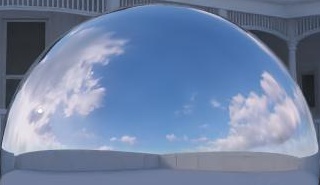} &
    \includegraphics[scale=0.4]{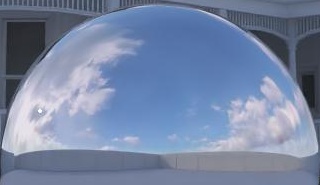} &
    \includegraphics[scale=0.4]{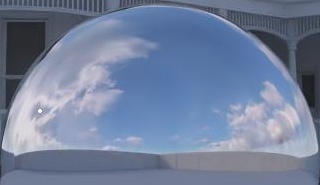} \\
    \rotatebox[origin=c]{90}{JazzyPicnic} &
    \includegraphics[scale=0.1]{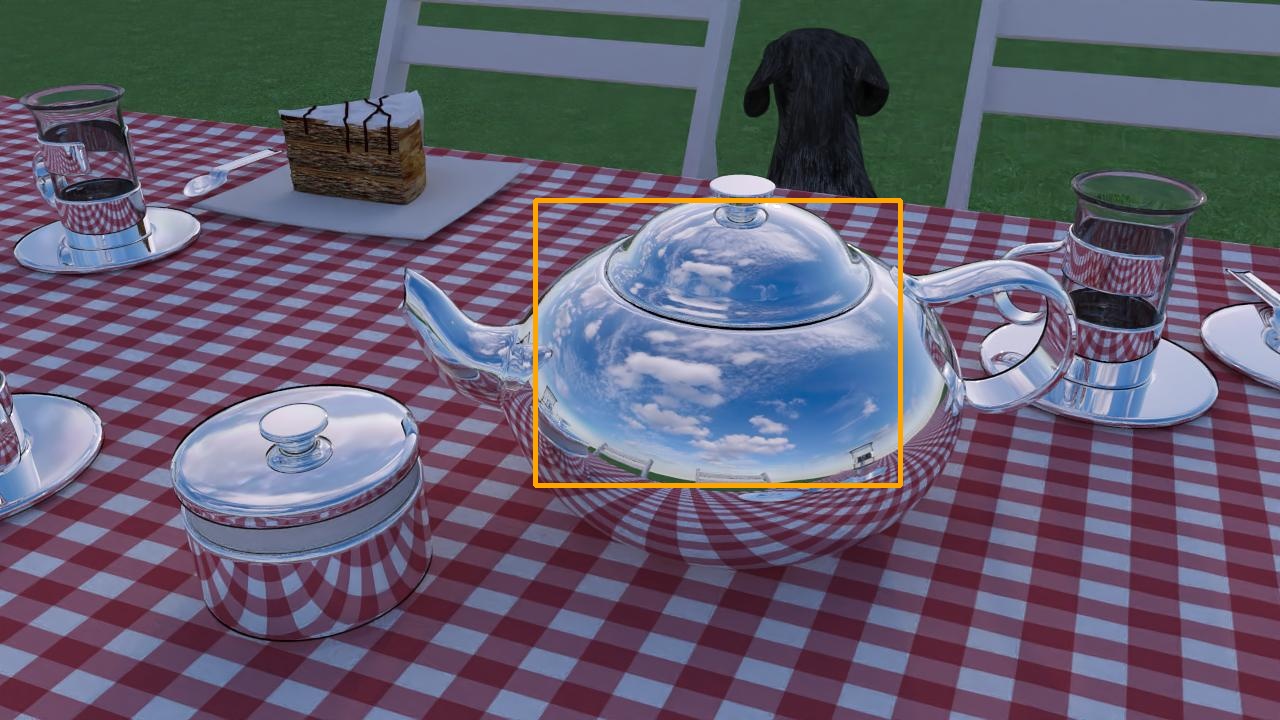} &
    \includegraphics[scale=0.1]{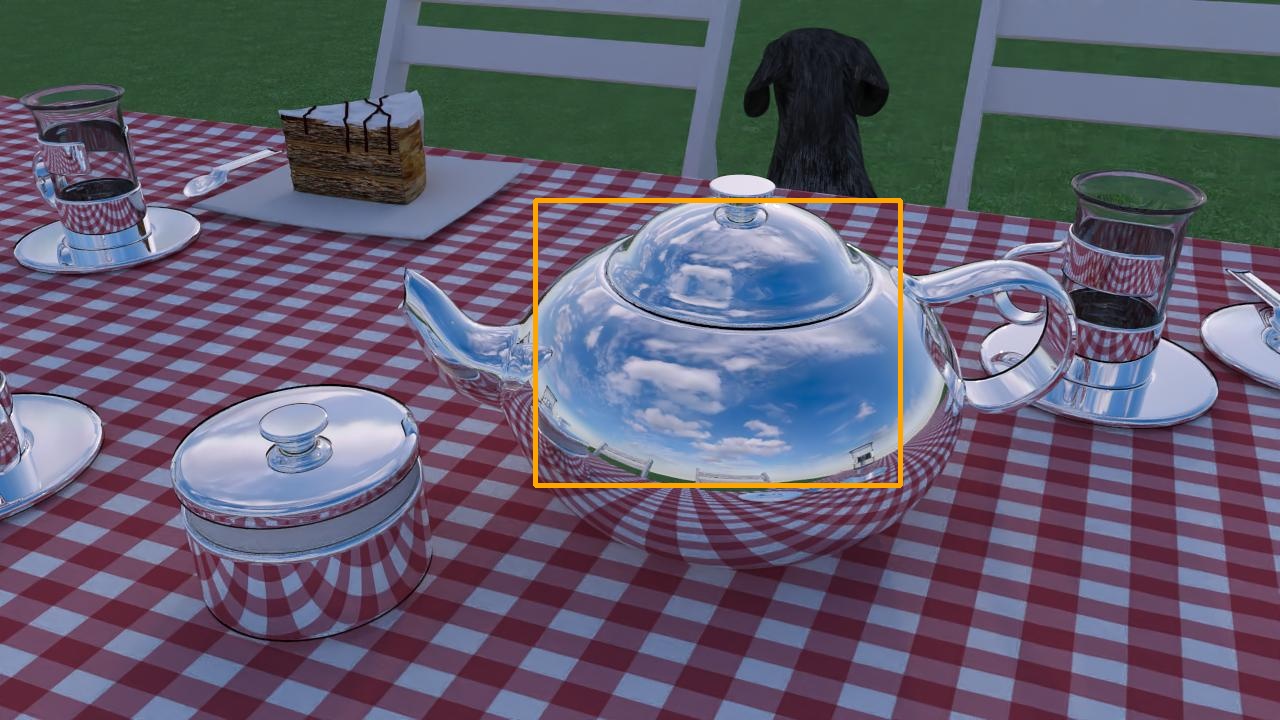} &
    \includegraphics[scale=0.1]{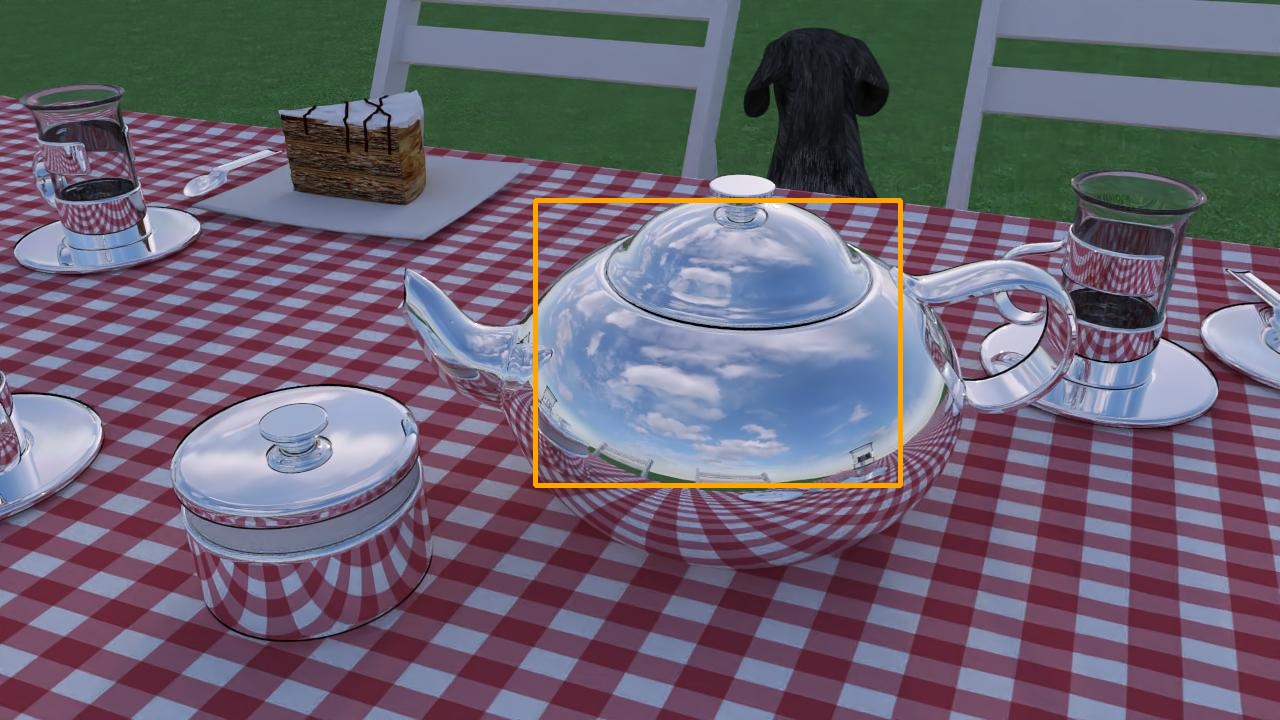} &
    \includegraphics[scale=0.1]{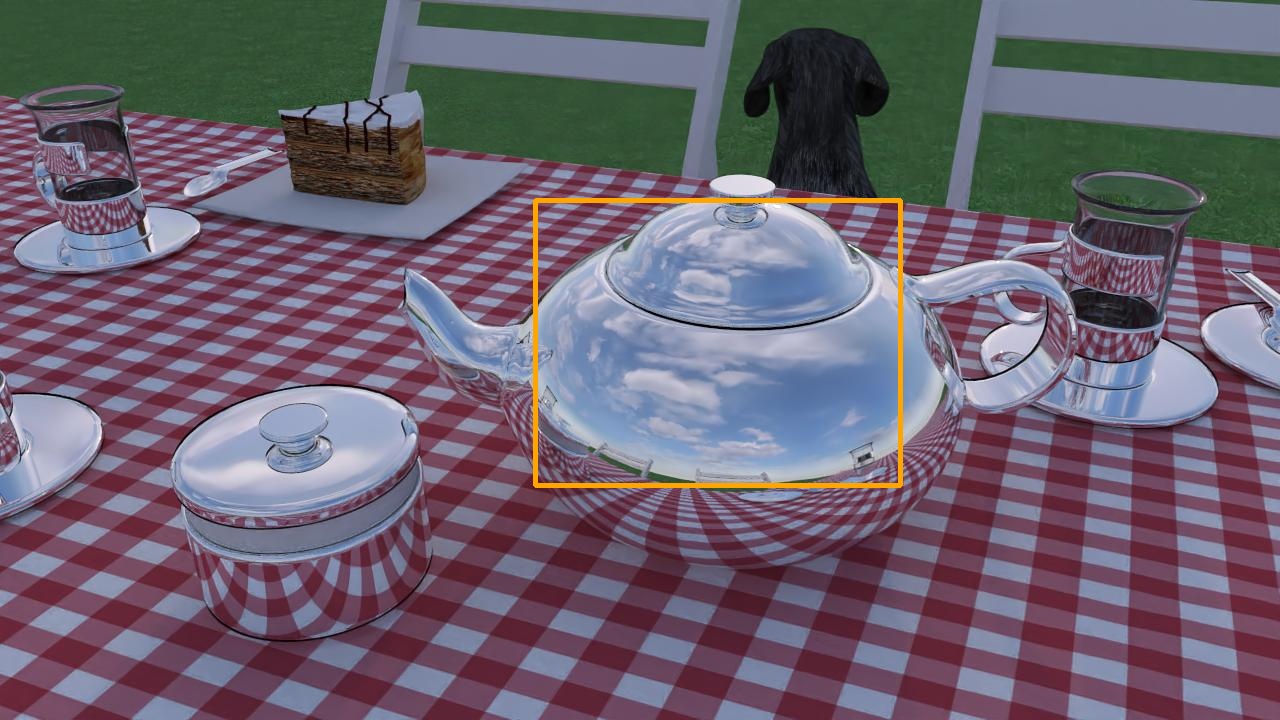} \\ [-5ex]
    & 
    \includegraphics[scale=0.35]{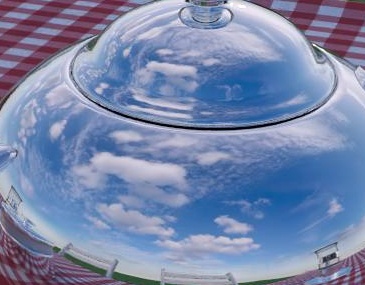} &
    \includegraphics[scale=0.35]{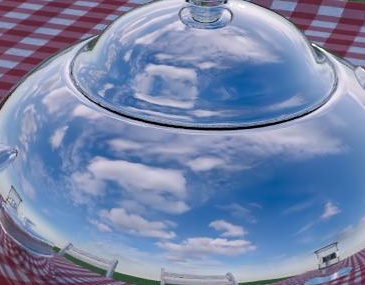} &
    \includegraphics[scale=0.35]{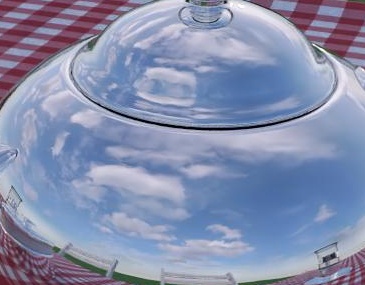} &
    \includegraphics[scale=0.35]{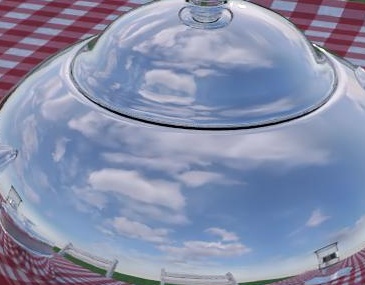} \\
    \rotatebox[origin=c]{90}{Observatory} & 
    \includegraphics[scale=0.1]{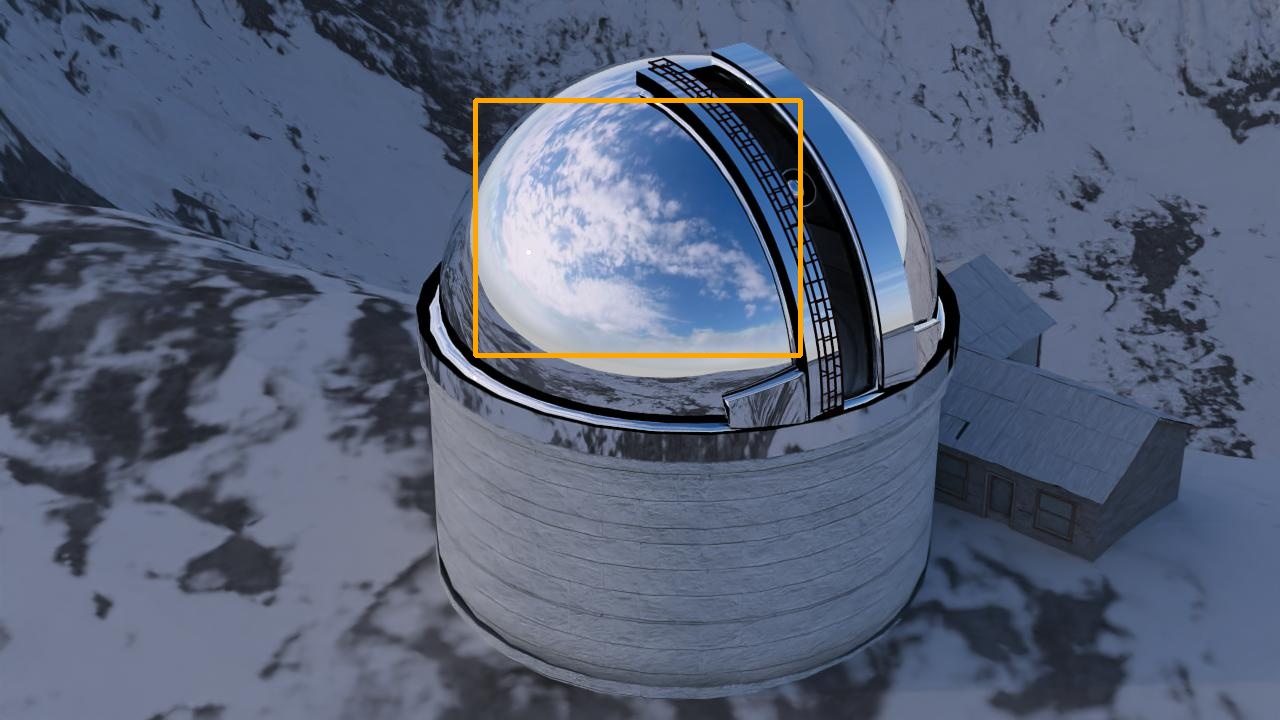} &
    \includegraphics[scale=0.1]{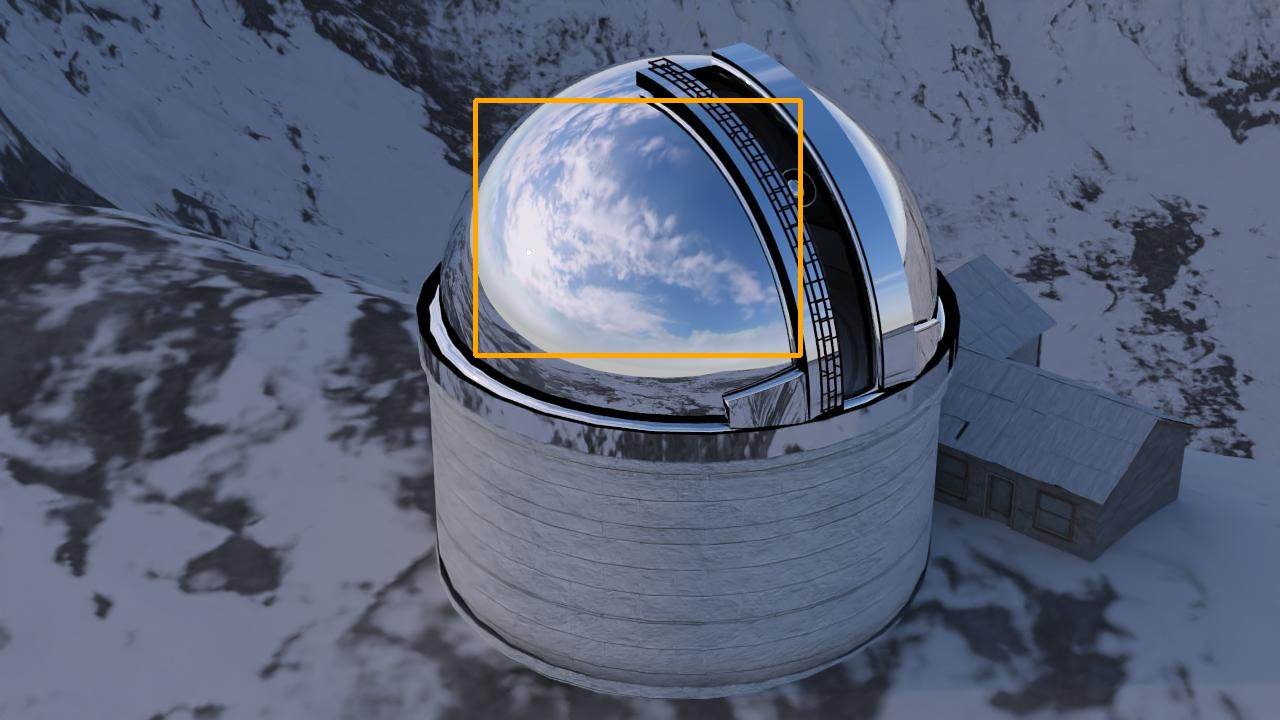} &
    \includegraphics[scale=0.1]{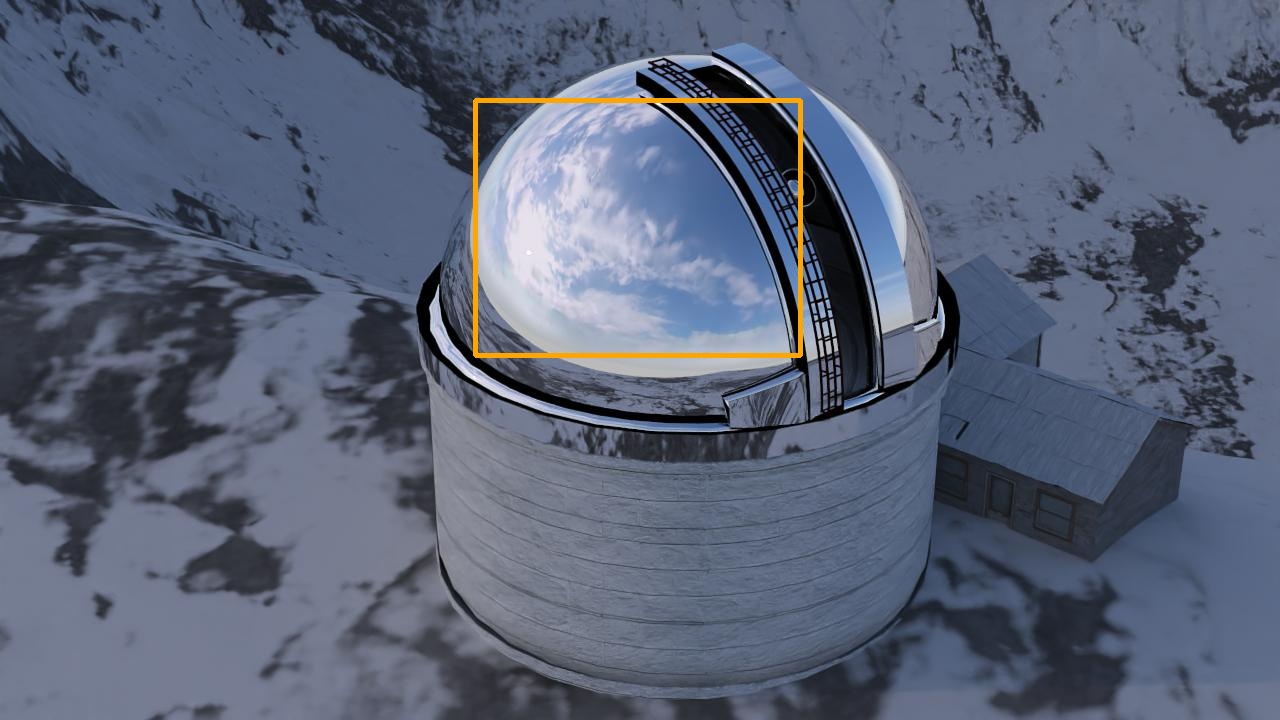} &
    \includegraphics[scale=0.1]{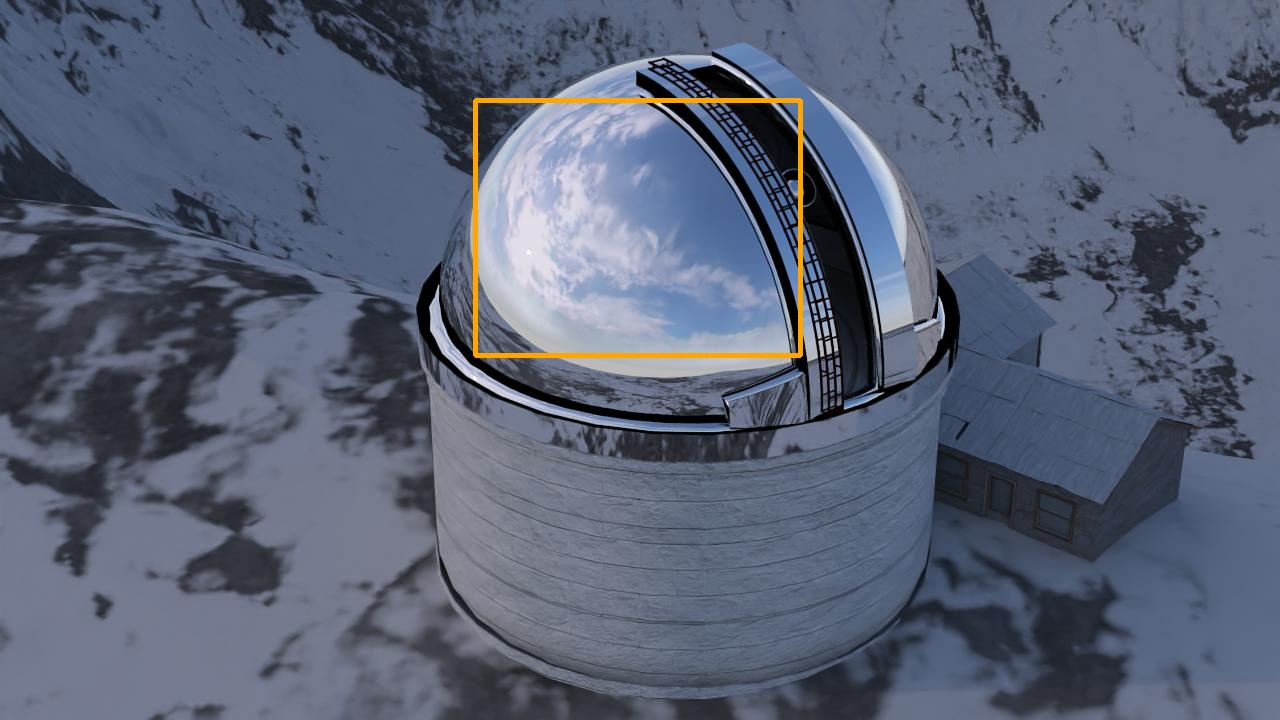} \\ [-5ex]
    & 
    \includegraphics[scale=0.393]{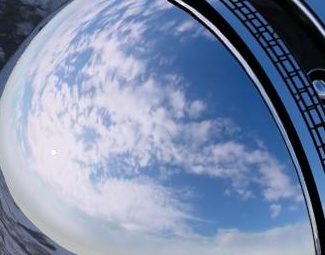} &
    \includegraphics[scale=0.393]{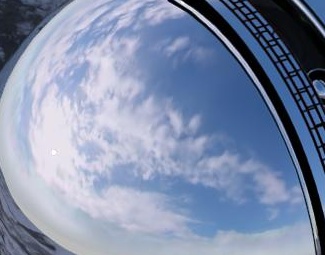} &
    \includegraphics[scale=0.393]{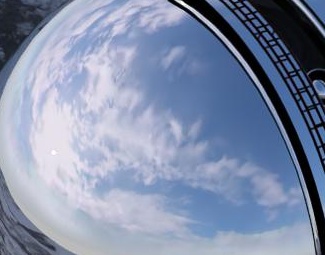} &
    \includegraphics[scale=0.393]{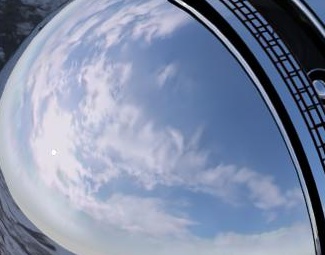}
\end{tabular}
\caption{Renderings showing our method being used to light a variety of scenes with different clouds over time. These frames are captured every 45 seconds and show our method can synthesize plausible cloud appearance over time.}
\label{fig:res1}
\end{figure*}

    We also show results for sky models being used for lighting virtual environments which is their expected use case. This shows that our method can produce sky imagery at a high enough resolution ($2048 \times 2048$ pixels) to be used for practical rendering purposes. Figure \ref{fig:res1} shows stills from a variety of 3D scenes showing a direct view of part of the sky (``Tower" and ``House" scenes), or reflections of a significant portion of the sky hemisphere (the ``House", ``JazzyPicnic" and ``Observatory" scenes) being illuminated with the dynamic cloud model proposed in this work (the lighting is computed using a path tracer). Please see the supplementary material for the full videos.

Figure \ref{fig:generatedres} shows frames of an animation using the output of a generative cloudy sky model in the ``JazzyPicnic", ``House" and ``Trophy Stadium" scenes, corresponding to the cloud masks shown in Figure \ref{fig:prevgenerated}. This again shows the realistic movement of clouds in a 3D environment, including when the initial clouds are artist controlled.

\begin{figure*}[htp]
\begin{tabular}{@{}l@{}l@{}l@{}l@{}l}
\rotatebox[origin=c]{90}{JazzyPicnic} &
    \includegraphics[scale=0.1]{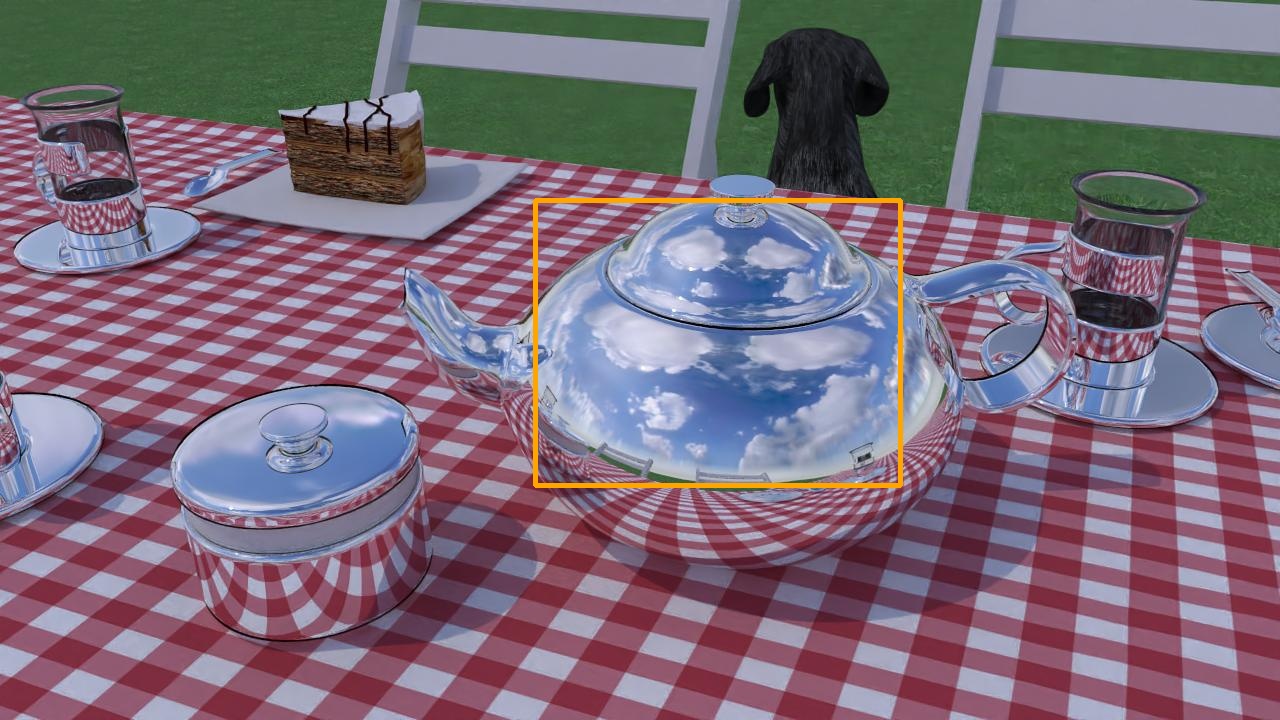} &
    \includegraphics[scale=0.1]{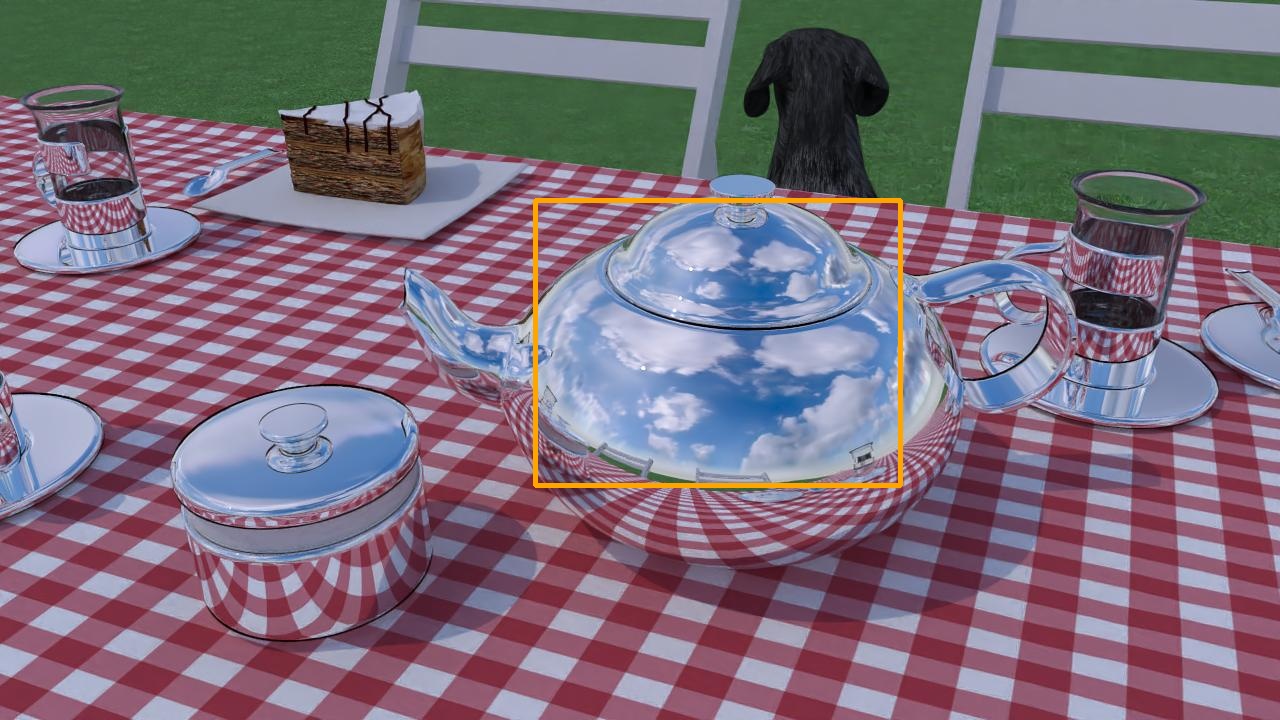} &
    \includegraphics[scale=0.1]{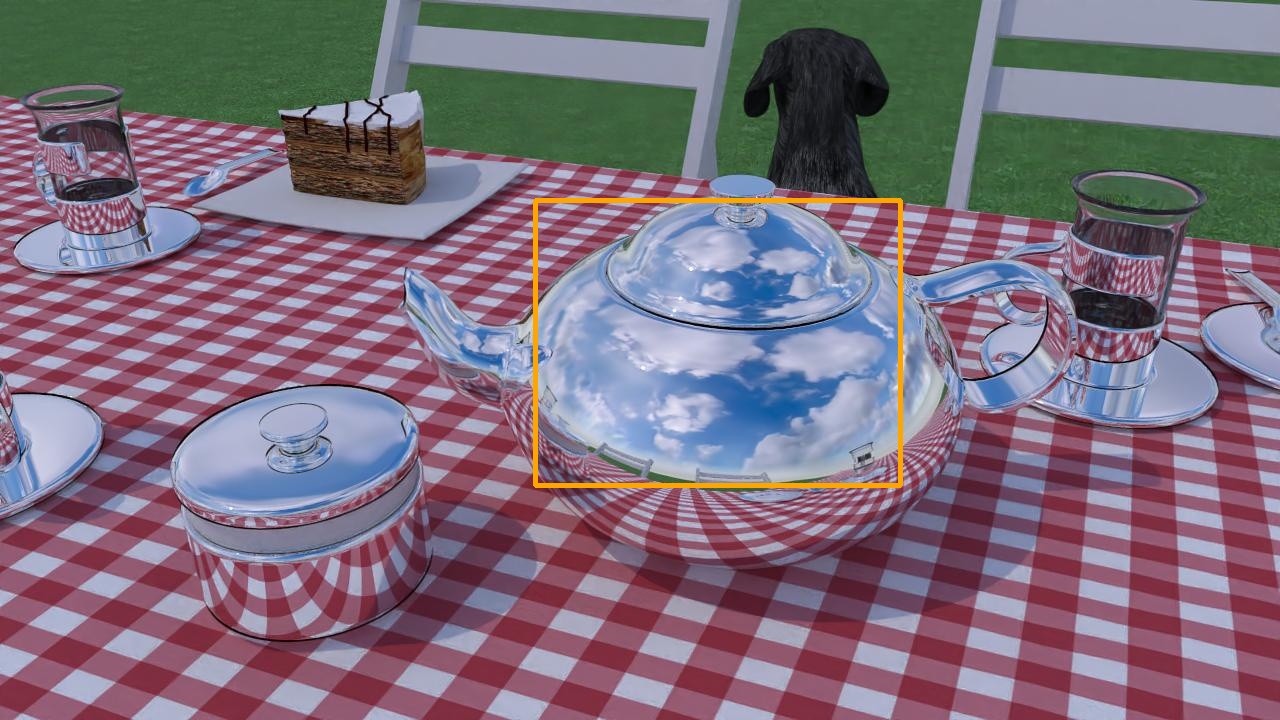} &
    \includegraphics[scale=0.1]{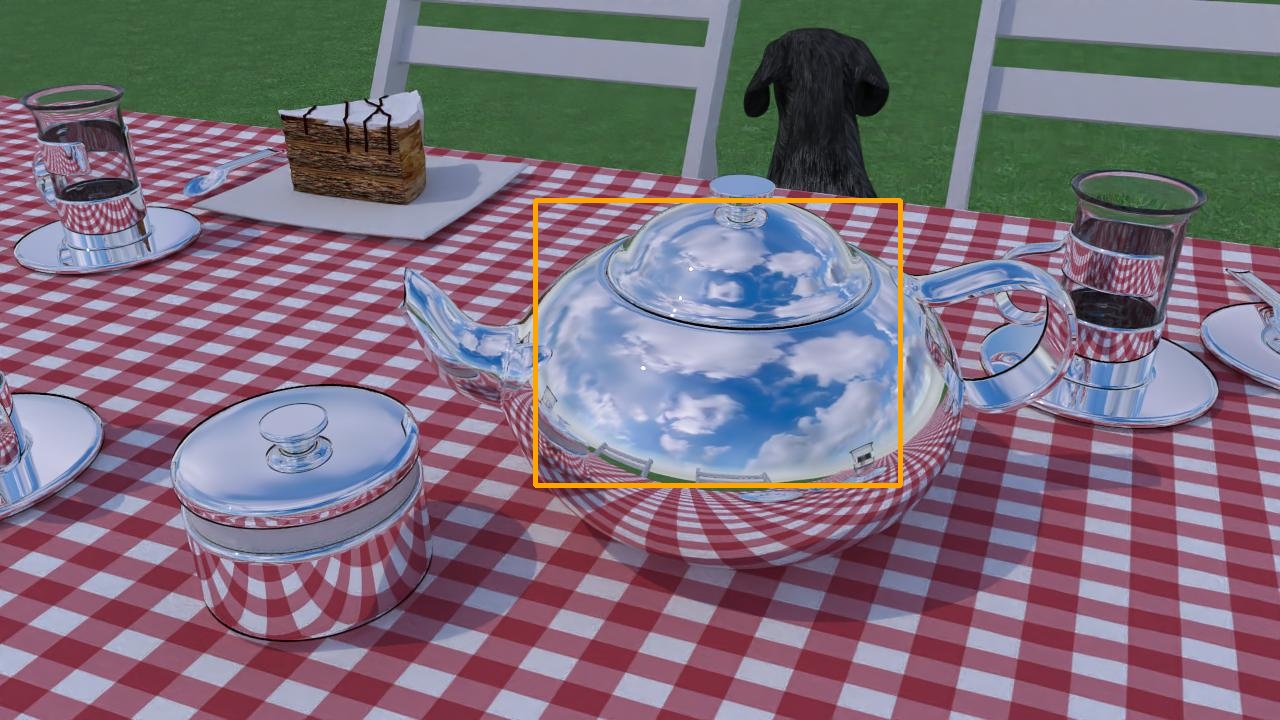} \\ [-5.5ex]
    & 
    \includegraphics[scale=0.351]{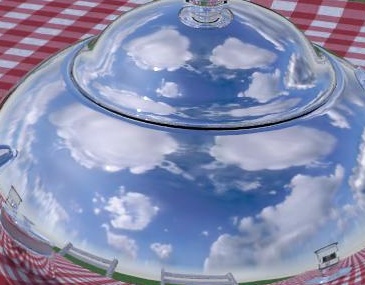} &
    \includegraphics[scale=0.351]{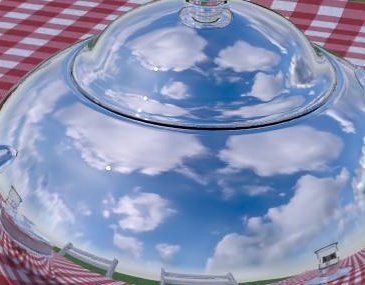} &
    \includegraphics[scale=0.351]{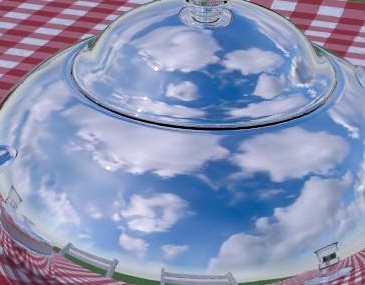} &
    \includegraphics[scale=0.351]{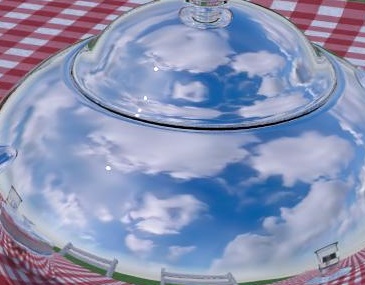} \\
\rotatebox[origin=c]{90}{House} &
    \includegraphics[scale=0.1]{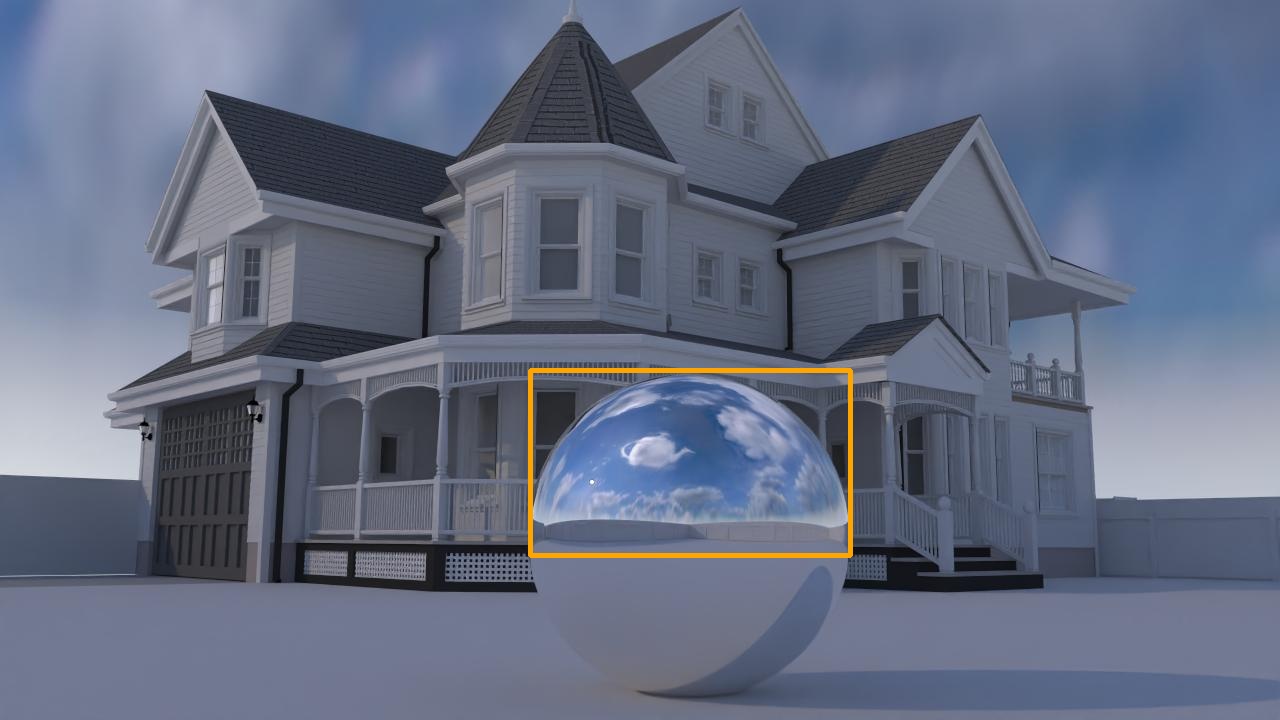} &
    \includegraphics[scale=0.1]{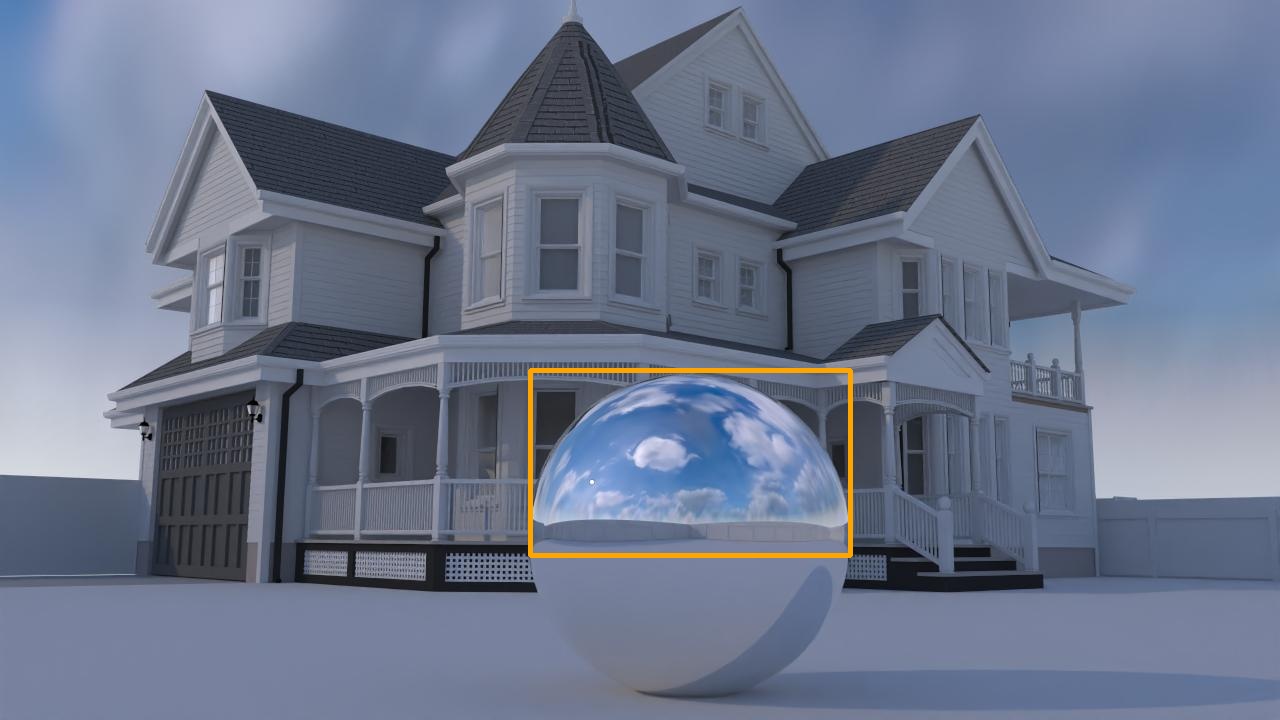} &
    \includegraphics[scale=0.1]{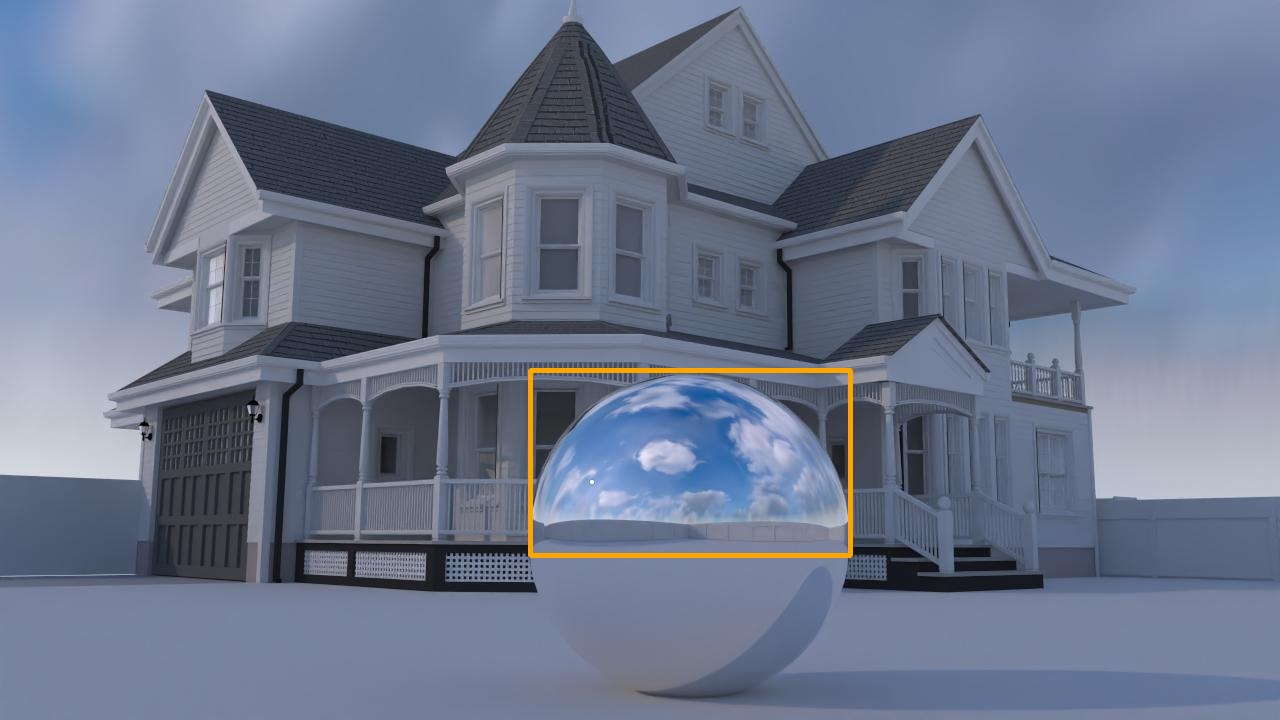} &
    \includegraphics[scale=0.1]{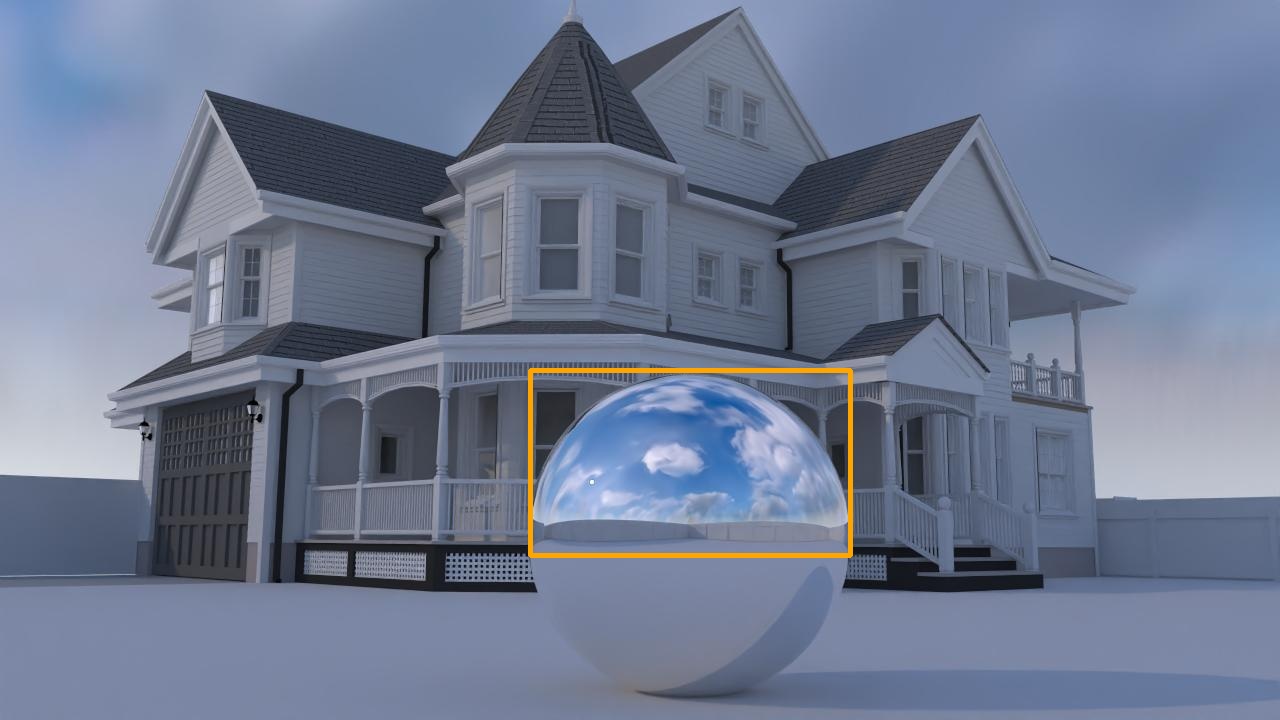} \\ [-5ex]
    & 
    \includegraphics[scale=0.4]{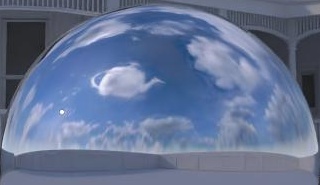} &
    \includegraphics[scale=0.4]{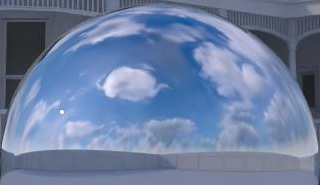} &
    \includegraphics[scale=0.4]{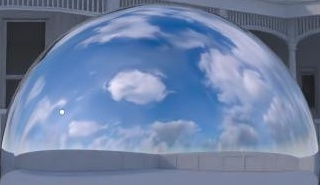} &
    \includegraphics[scale=0.4]{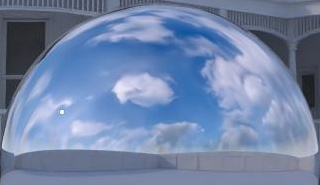} \\ 
\rotatebox[origin=c]{90}{Trophy Stadium} &
    \includegraphics[scale=0.1]{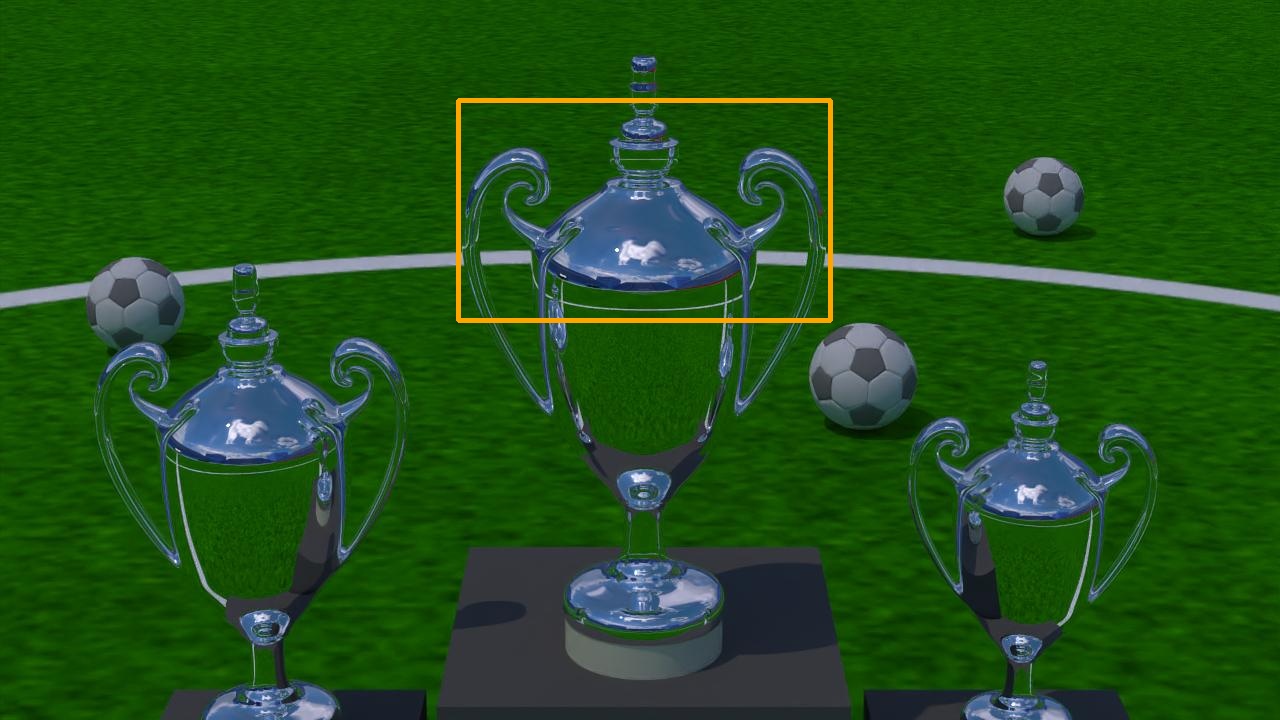} &
    \includegraphics[scale=0.1]{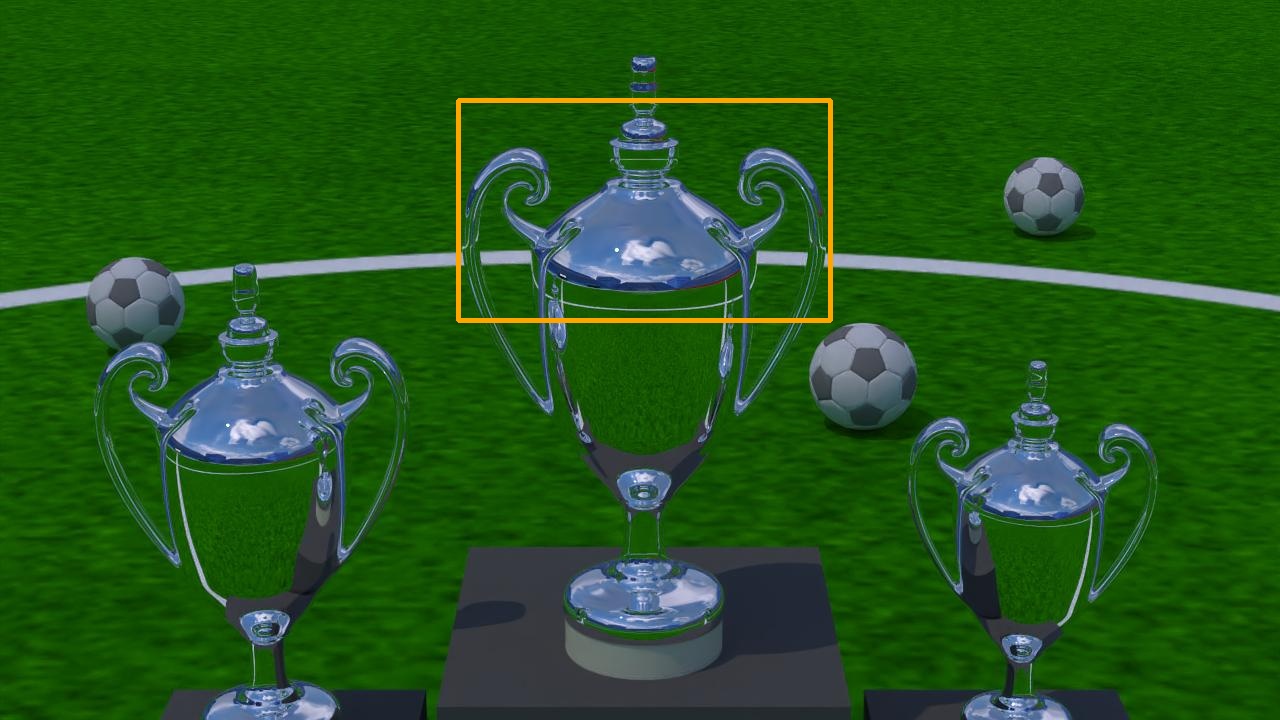} &
    \includegraphics[scale=0.1]{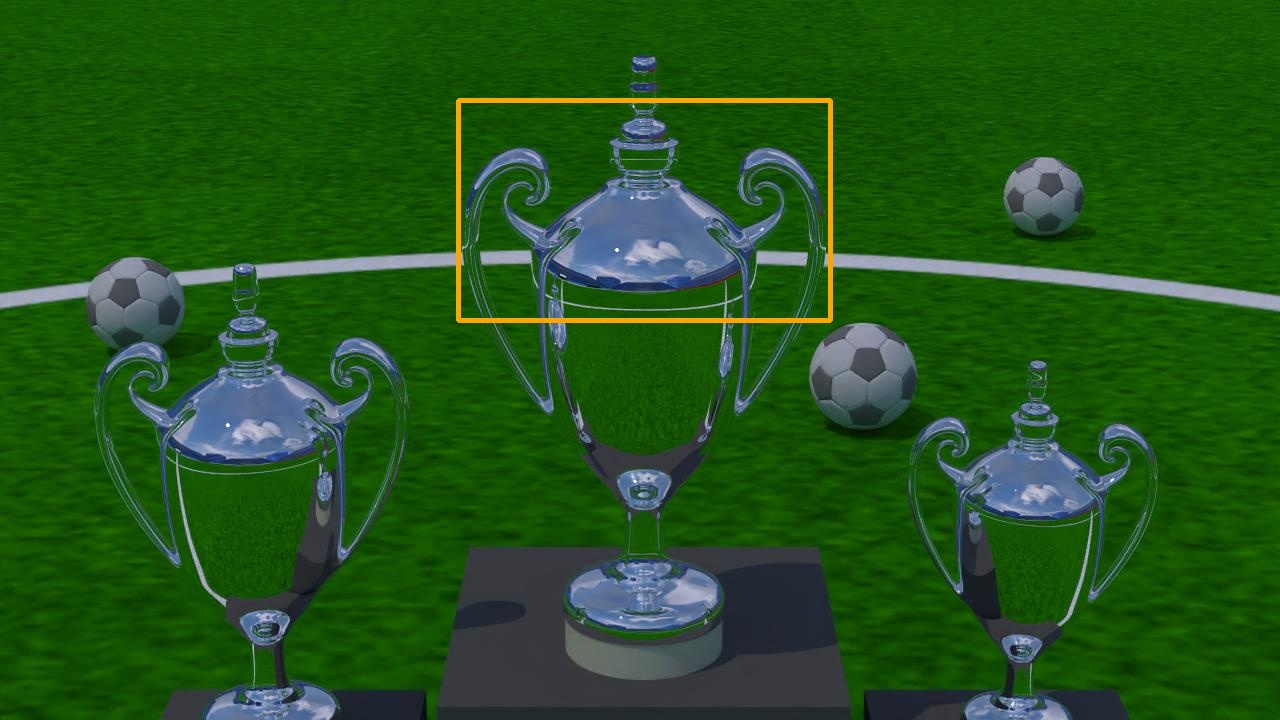} &
    \includegraphics[scale=0.1]{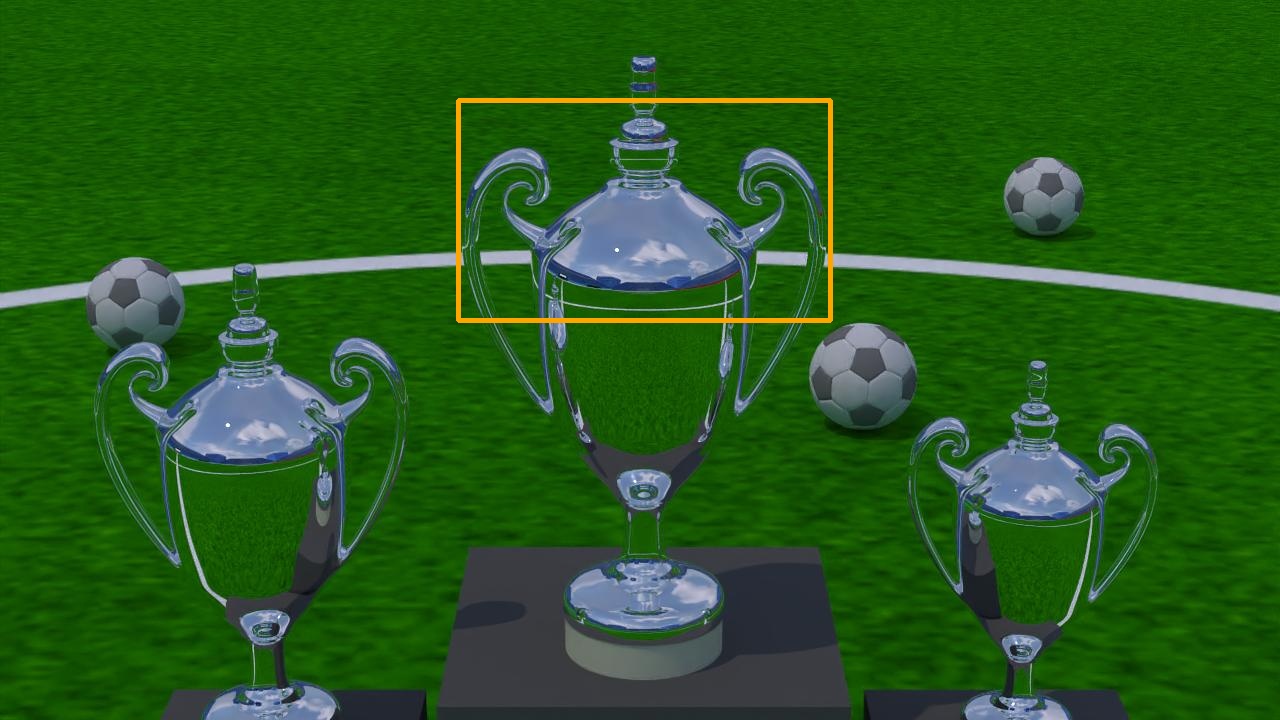} \\ [-6.75ex]
    & 
    \includegraphics[scale=0.343516]{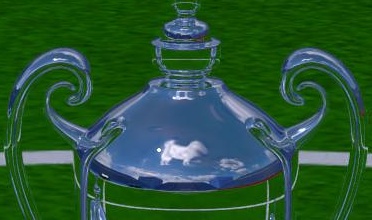} &
    \includegraphics[scale=0.343516]{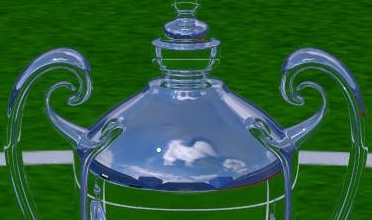} &
    \includegraphics[scale=0.343516]{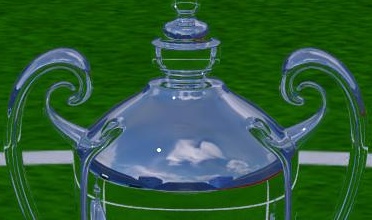} &
    \includegraphics[scale=0.343516]{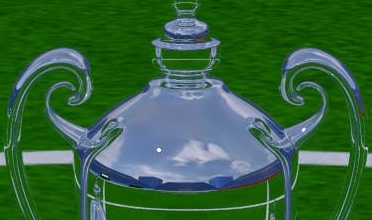} \\
\end{tabular}
\caption{Renderings showing the use of outputs from deep synthesis model, in this case showing clouds generated from a mask of a ``Thought Bubbles'' (top), Teapot (middle), and Dog (bottom). Our method provides photorealistic cloud animation even in this very artificial scenario.}
\label{fig:generatedres}
\end{figure*}

\subsection{Impact of loss functions}

\begin{table}[!h]
\centering
\begin{tabular}{|l|l|l|l|l|}
\hline
      & MSE     & MSE+Cos  \\ \hline  % & Cos   
Error & 0.0085  & 0.0009   \\ \hline  % & 1.0812 
\end{tabular} \vspace{1mm} 
\caption{Ablation study for loss functions used in this work. This shows that using MSE + Cosine Similarity leads to substantially less error than MSE.}
\label{table:ablation}
\end{table}

We also conducted an ablation study to assess the use of MSE verses MSE+Cosine loss functions for training the network in Section \ref{subsec:NonLinearModel}. The results can be seen in Table \ref{table:ablation}. The results are obtained through training 500 epochs of $FlowNet$ and $CloudNet$. It can be seen that using the combination of MSE and Cosine loss functions together provides substantially higher accuracy than only using MSE.

\section{Conclusion}

This work has proposed a multi-timescale sky appearance model which adds dynamic cloud movement to static hemispherical cloud images. We show that our approach can synthesize plausible cloud movement for a wide range of captured skies and can also add dynamism to prior work which generates static sky imagery. We achieved this via a multi-scale approach which uses neural networks and flow fields to predict sky illumination at fixed intervals, and proposed a principled interpolation approach at short time-scales.

In the future, we aim to extend our approach to integrate longer term weather forecasting, possibly via graph neural networks combined with generative models, and to extend our dataset to include rarer cloud types. Finally, we are interested in combining our image based method with light transport methods to additionally synthesize 3D volumetric clouds.

\section{Acknowledgements}

We gratefully acknowledge the support of NVIDIA Corporation with the donation of the RTX A6000 used for this research. We also would like to thank the following for the scenes and models used in this work: MrChimp2313, Alim Zhilov, Noker, timothyjamesbusch, alplaleli, and Helena-Merlot.

\bibliographystyle{IEEEtran}
\bibliography{refs}

\end{document}